%% file: main.tex
\documentclass[camera]{jpaper}
\usepackage{makecell}

\usepackage[nocompress]{cite}

\usepackage{soul}

\usepackage{amsmath}
\usepackage{graphicx}
\usepackage[linesnumbered,ruled]{algorithm2e}

\usepackage{algpseudocode}
\usepackage{titlesec}
\algrenewcommand\algorithmiccomment[2][\normalsize]{{#1\hfill\(\triangleright\) #2}}
\usepackage{geometry} \titlespacing*{\section}{0pt}{2pt plus 0.5pt minus 0.5pt}{0pt}
\titlespacing*{\subsection}{0pt}{2pt plus 0.5pt minus 0.5pt}{0pt}
\titlespacing*{\subsubsection}{0pt}{2pt plus 0.5pt minus 0.5pt}{0pt}

\makeatletter
\renewcommand{\@makefnmark}{\hbox{\textsuperscript{\scriptsize{\@thefnmark}}}}
\makeatother

\usepackage{array}
\newcolumntype{L}[1]{>{\raggedright\let\newline\\\arraybackslash\hspace{0pt}}m{#1}}
\newcolumntype{C}[1]{>{\centering\let\newline\\\arraybackslash\hspace{0pt}}m{#1}}
\newcolumntype{R}[1]{>{\raggedleft\let\newline\\\arraybackslash\hspace{0pt}}m{#1}}

\usepackage{bm}

\makeatletter
\let\MYcaption\@makecaption
\makeatother

\usepackage[font=footnotesize]{subcaption}

\makeatletter
\let\@makecaption\MYcaption
\makeatother

\usepackage{fixltx2e}

\usepackage{dblfloatfix}
\usepackage[nolessnomore, italic]{mathastext}
\usepackage[T1]{fontenc}
\usepackage[usenames,dvipsnames,svgnames,table]{xcolor}
\usepackage{multirow}
\usepackage{hhline}
\usepackage[normalem]{ulem}
\usepackage{setspace}
\usepackage{indentfirst}
\usepackage{footmisc}

\usepackage{pifont}

\newcommand{\stripe}{\rowcolor{blue!5}}
\newcommand{\mr}[2]{\multicolumn{1}{c}{\multirow{#1}{*}{\makecell{#2}}}}

\newif\ifcameraready
\camerareadyfalse

\ifcameraready

\newcommand{\affilCMU}[0]{\textsuperscript{$\ddagger$}}
\else

\newcommand{\affilCMU}[0]{\textsuperscript{$\dagger$}}
\fi
\newcommand{\affilETH}[0]{\textsuperscript{\S}}

\definecolor{amber}{rgb}{1.0, 0.49, 0.0}
\definecolor{darkgreen}{rgb}{0.0, 0.2, 0.13}
\definecolor{darkbyzantium}{rgb}{0.36, 0.22, 0.33}
\definecolor{darkseagreen}{rgb}{0.56, 0.74, 0.56}
\definecolor{darkspringgreen}{rgb}{0.09, 0.45, 0.27}
\definecolor{dollarbill}{rgb}{0.52, 0.73, 0.4}

\newcommand{\hcfirst}[0]{$HC_{\scalebox{0.7}{first}}$}  
\newcommand{\hcsecond}[0]{$HC_{\scalebox{0.7}{second}}$}  
\newcommand{\hcthird}[0]{$HC_{\scalebox{0.7}{third}}$}  

\newcommand{\ap}[1]{{#1}}

\newcommand{\delete}[1]{{\color{red}\sout{}}}

\newcommand{\versionnum}[0]{9.0\today~@ 17:40 CET} 
\newcommand{\microsubmissionnumber}{137}

\fancypagestyle{firstpage}{
  
  \fancyhead[C]{\normalsize{HPCA 2019 Submission
  \textbf{\#\microsubmissionnumber} -- Confidential Draft -- Do NOT Distribute!! \\
  \textcolor{MidnightBlue}{\emph{Version \versionnum}}
  }}
  \pagenumbering{arabic}
}

\begin{document}
\bstctlcite{IEEEexample:BSTcontrol} 


\title{\vspace{-5mm}Revisiting RowHammer: An Experimental Analysis \\ of Modern DRAM Devices and Mitigation Techniques}


%


\author{
{Jeremie S. Kim\affilETH\affilCMU}\qquad%
{Minesh Patel\affilETH}\qquad%
{A. Giray Ya\u{g}l{\i}k\c{c}{\i}\affilETH}\qquad \\
{Hasan Hassan\affilETH}\qquad%
{Roknoddin Azizi\affilETH}\qquad 
{Lois Orosa\affilETH}\qquad%
{Onur Mutlu\affilETH\affilCMU}\qquad\vspace{-3mm}\\\\%
{\vspace{-3mm}\emph{\affilETH ETH Z{\"u}rich \qquad \affilCMU Carnegie Mellon University}}%
}


%


\maketitle
\thispagestyle{plain} 
\pagestyle{plain}

\setstretch{0.8}
\renewcommand{\footnotelayout}{\setstretch{0.9}}

\camerareadytrue


\fancyhead{}
\ifcameraready
 \thispagestyle{plain}
 \pagestyle{plain}
\else
 \fancyhead[C]{\textcolor{MidnightBlue}{\emph{Version \versionnum~---~}}}
 \fancypagestyle{firststyle}
 {
   \fancyhead[C]{\textcolor{MidnightBlue}{\emph{Version \versionnum~---~}}}
   \fancyfoot[C]{\thepage}
 }
 \thispagestyle{firststyle}
 \pagestyle{firststyle}
\fi




%


\input{0_abstract}

\input{1_introduction}
\input{2_background}
\input{3_motivation}

\input{4_methodology}
\input{5_characterization}
\input{6_implications_jk}
\input{7_related}
\input{8_conclusion}


\section*{Acknowledgments} We thank the anonymous reviewers of ISCA 2020 for
feedback and the SAFARI group members for feedback and the stimulating
intellectual environment they provide. We also thank our industrial
partners for their generous donations.



%


\SetTracking
 [ no ligatures = {f},
 outer kerning = {*,*} ]
 { encoding = * }
 { -40 } 

{

  \footnotesize
  \renewcommand{\baselinestretch}{0.5}
  \let\OLDthebibliography\thebibliography
  \renewcommand\thebibliography[1]{
    \OLDthebibliography{#1}
    \setlength{\parskip}{0pt}
    \setlength{\itemsep}{0pt}
  }
  \bibliographystyle{IEEEtranS}
  \bibliography{ref}
}

\newpage 
\input{appendix}

\end{document}

%% file: 0_abstract.tex
\begin{abstract} 
RowHammer is a circuit-level DRAM vulnerability, first rigorously analyzed
and introduced in 2014, where repeatedly accessing data in a DRAM row can cause
bit flips in nearby rows. The RowHammer vulnerability has since garnered
significant interest in both computer architecture and computer security
research communities because it stems from physical circuit-level interference
effects that worsen with continued DRAM density scaling. As
DRAM manufacturers primarily depend on density scaling to increase DRAM
capacity, future DRAM chips will likely be more vulnerable to RowHammer than
those of the past. Many RowHammer mitigation mechanisms have been proposed by
both industry and academia, but it is unclear whether these mechanisms will
remain viable solutions for future devices, as their overheads increase
with DRAM's vulnerability to RowHammer. 

In order to shed more light on how RowHammer affects modern and future devices
at the circuit-level, we first present an experimental characterization of
RowHammer on 1580 DRAM chips (408$\times$ DDR3, 652$\times$ DDR4, and
520$\times$ LPDDR4) from 300 DRAM modules (60$\times$ DDR3, 110$\times$
DDR4, and 130$\times$ LPDDR4) with RowHammer protection mechanisms disabled,
spanning multiple different technology nodes from across each of the three
major DRAM manufacturers.  Our studies definitively show that newer DRAM chips
are more vulnerable to RowHammer: as device feature size reduces, the
number of activations needed to induce a RowHammer bit flip also reduces, to as
few as $9.6k$ (4.8k to two rows each) in the most vulnerable chip we tested.  

We evaluate five state-of-the-art RowHammer mitigation mechanisms using
cycle-accurate simulation in the context of real data taken from our chips to
study how the mitigation mechanisms scale with chip vulnerability. We find
that existing mechanisms either are not scalable or suffer from prohibitively
large performance overheads in projected future devices
given our observed trends of RowHammer vulnerability. Thus, it is critical to
research more effective solutions to RowHammer. 

\end{abstract}

%% file: 1_introduction.tex
\section{Introduction}
\label{sec:introduction}

DRAM is the dominant main memory technology of nearly all modern
computing systems due to its superior cost-per-capacity. As such, DRAM
critically affects overall system performance and reliability.
Continuing to increase DRAM capacity requires increasing the density of
DRAM cells by reducing (i.e., scaling) the technology node size (e.g.,
feature size) of DRAM, but this scaling negatively impacts DRAM reliability. In
particular, RowHammer~\cite{kim2014flipping} is an important circuit-level
interference phenomenon, closely related to technology scaling, where
repeatedly activating a DRAM row disturbs the values in adjacent rows.
RowHammer can result in system-visible bit flips in DRAM regions that are
\emph{physically nearby} rapidly accessed (i.e., hammered) DRAM rows.
RowHammer empowers an attacker who has access to DRAM address $X$ with
the ability to modify data in a different location $Y$ such that $X$ and $Y$
are \emph{physically}, but not necessarily \emph{logically}, co-located. In
particular, $X$ and $Y$ must be located in different DRAM rows that are in
close proximity to one another. Because DRAM is widely used throughout
modern computing systems, many systems are potentially vulnerable to RowHammer
attacks, as shown by recent works (e.g., \cite{cojocar2019exploiting,
gruss2018another, gruss2016rowhammer, lipp2018nethammer, qiao2016new,
razavi2016flip, seaborn2015exploiting, tatar2018defeating, van2016drammer,
xiao2016one, frigo2020trrespass, ji2019pinpoint, kwong2020rambleed,
mutlu2017rowhammer, van2018guardion}).  

RowHammer is a serious challenge for system designers because it exploits
fundamental DRAM circuit behavior that cannot be easily changed. This means
that RowHammer is a potential threat across all DRAM generations and designs.
Kim et al.~\cite{kim2014flipping} show that RowHammer appears to be an effect
of continued DRAM technology scaling~\cite{kim2014flipping, mutlu2013memory,
mutlu2019rowhammer, mutlu2019rowhammerbeyond}, which means that as
manufacturers increase DRAM storage density, their chips are potentially more
susceptible to RowHammer.  This increase in RowHammer vulnerability is often
quantified for a given DRAM chip by measuring the number of times a single row
must be activated (i.e., single-sided RowHammer) to induce the first bit
flip.  Recently, Yang et al.~\cite{yang2019trap} have corroborated this
hypothesis, identifying a precise circuit-level charge leakage mechanism that
may be responsible for RowHammer. This leakage mechanism affects nearby circuit
components, which implies that as manufacturers continue to employ aggressive
technology scaling for generational storage density
improvements~\cite{mandelman2002challenges, kang2014co,
vogelsang2010understanding, hong2010memory}, circuit components that are more
tightly packed will likely increase a chip's vulnerability to RowHammer.


To mitigate the impact of the RowHammer problem, numerous works propose
\emph{mitigation mechanisms} that seek to prevent RowHammer bit flips from
affecting the system. These include mechanisms to make RowHammer
conditions impossible or very difficult to attain (e.g., increasing the default
DRAM refresh rate by more than 7x~\cite{kim2014flipping}, or probabilistically
activating adjacent rows with a carefully selected
probability~\cite{kim2014flipping}) and mechanisms that explicitly detect
RowHammer conditions and intervene (e.g., access counter-based
approaches~\cite{lee2019twice, seyedzadeh2017counter, seyedzadeh2018cbt,
kim2014flipping, kim2014architectural}). However, \emph{all} of these
solutions~\cite{you2019mrloc, son2017making, aweke2016anvil, konoth2018zebram,
van2018guardion, brasser2016can, kim2014flipping, kim2014architectural,
irazoqui2016mascat, gomez2016dram, lee2019twice, bu2018srasa, bains2015rowref,
bains14d, bains14c, greenfield14b, bains2016row, bains2015row, rh-apple, rh-hp,
rh-lenovo, rh-cisco, hassan2019crow, wu2019protecting, bock2019rip,
kim2019effective, wang2019reinforce, fisch2017dram, chakraborty2019deep,
li2019detecting, wang2019detect, ghasempour2015armor} merely treat the symptoms
of a RowHammer attack (i.e., prevent RowHammer conditions)
without solving the core circuit vulnerability.

To better understand the problem in order to pursue more comprehensive
solutions, prior works study the RowHammer failure mechanism both
experimentally~\cite{kim2014flipping, park2016statistical,
park2016experiments} and in simulation~\cite{yang2019trap}. Unfortunately,
there has been no work since the original RowHammer
paper~\cite{kim2014flipping} that provides a rigorous characterization-based
study to demonstrate how chips' vulnerabilities to RowHammer (i.e.,
the minimum number of activations required to induce the first RowHammer bit
flip) scale across different DRAM technology generations. While many
works~\cite{mutlu2019rowhammerbeyond, bains2015row, kim2016rowhammer,
mutlu2017rowhammer, mutlu2019rowhammer} speculate that modern chips are more
vulnerable, there is no rigorous experimental study that demonstrates exactly
how the minimum activation count to induce the first RowHammer bit flip and
other RowHammer characteristics behave in modern DRAM chips. Such an
experimental study would enable us to predict future chips' vulnerability to
RowHammer and estimate whether existing RowHammer mitigation mechanisms
can effectively prevent RowHammer bit flips in modern and future chips. 

\textbf{Our goal} in this work is to experimentally demonstrate how vulnerable
modern DRAM chips are to RowHammer at the circuit-level and to study how this
vulnerability will scale going forward. To this end, we provide a rigorous
experimental characterization of 1580 DRAM chips (408$\times$ DDR3, 652$\times$
DDR4, and 520$\times$ LPDDR4) from 300 modern DRAM modules (60$\times$ DDR3,
110$\times$ DDR4, and 130$\times$ LPDDR4) from across all three major
DRAM manufacturers, spanning across multiple different technology node
generations for each manufacturer. To study the RowHammer vulnerability at the
\emph{circuit} level instead of at the \emph{system} level, we disable all accessible RowHammer
mitigation mechanisms.\footnote{We cannot disable on-die ECC in our LPDDR4
chips~\cite{micron2017whitepaper, kwak2017a, kang2014co,
patel2019understanding, kwon2017an}.} We observe that the worst-case
circuit-level RowHammer conditions for a victim row are when we repeatedly
access \emph{both} physically-adjacent aggressor rows as rapidly as possible
(i.e., \emph{double-sided RowHammer}). To account for double-sided RowHammer in
our study, we define \emph{Hammer Count} ($HC$) as the number of times \emph{each} 
physically-adjacent row is activated and {\hcfirst} as the minimum $HC$
required to cause the first RowHammer bit flip in the DRAM chip. For each DRAM
type (i.e., DDR3, DDR4, LPDDR4), we have chips from at least two different
technology nodes, and for each of the LPDDR4 chips, we know the exact process
technology node: 1x or 1y. This enables us to study and demonstrate the effects
of RowHammer on two distinct independent variables: DRAM type and DRAM
technology node. 

Our experiments study the effects of manipulating two key testing
parameters at a fixed ambient temperature on both aggregate and individual DRAM
cell failure characteristics: (1) hammer count ($HC$), and (2) data pattern
written to DRAM. 

Our experimental results definitively show that newer DRAM chips manufactured
with smaller technology nodes are increasingly vulnerable to RowHammer bit
flips.  For example, we find that {\hcfirst} across all chips of a given DRAM type
reduces greatly from older chips to newer chips (e.g., $69.2k$ to $22.4k$ in
DDR3, $17.5k$ to $10k$ in DDR4, and $16.8k$ to $4.8k$ in LPDDR4
chips).\footnote{While we do not definitively know the exact technology nodes
used in our DDR3/DDR4 chips, we group our chips into two sets (i.e., new and
old) based on their manufacturing dates, datasheet publication dates,
purchase dates, and distinctive RowHammer characterization results. We compare
our results against those from our LPDDR4 chips whose exact technology nodes we
know and observe the same trend of higher RowHammer vulnerability
with newer chips (that likely use smaller DRAM process technology nodes).} 


Using the data from our experimental studies, we perform cycle-accurate
simulation to evaluate the performance overheads of five state-of-the-art
RowHammer mitigation mechanisms~\cite{kim2014flipping, lee2019twice,
son2017making, you2019mrloc} and compare them to an ideal refresh-based
RowHammer mitigation mechanism that selectively refreshes a row only just
before it is about to experience a RowHammer bit flip. We show that, while
the state-of-the-art mechanisms are reasonably effective at mitigating
RowHammer in today's DRAM chips (e.g., 8\% average performance loss in our
workloads when PARA~\cite{kim2014flipping} is used in a DRAM chip with an
{\hcfirst} value of $4.8k$), they exhibit prohibitively large performance
overheads for projected degrees of RowHammer vulnerability (i.e., lower
{\hcfirst} values) in future DRAM chips (e.g., the most-scalable existing
RowHammer mitigation mechanism causes 80\% performance loss when {\hcfirst}
is 128). This means that the state-of-the-art RowHammer mitigation mechanisms
are not scalable in the face of worsening RowHammer vulnerability, and
DRAM-based computing systems will either require stronger failure mitigation 
mechanisms or circuit-level modifications that address the root cause of the
RowHammer vulnerability. 

Our simulation results of an ideal refresh-based mitigation mechanism,
which selectively refreshes only those rows that are about to experience a
RowHammer bit flip, demonstrates significant opportunity for developing a
refresh-based RowHammer mitigation mechanism with low performance overhead that
scales reasonably to low {\hcfirst} values.  However, we observe that even this
ideal mechanism significantly impacts overall system performance at very
low {\hcfirst} values, indicating the potential need for a better approach to
solving RowHammer in the future. We discuss directions for future research
in this area in Section~\ref{subsec:implications:mitigation_solutions}. 




We make the following contributions in this work: 
\begin{itemize}
    \item We provide the first rigorous RowHammer failure characterization study of a broad range of real modern DRAM chips across different DRAM types, technology node generations, and manufacturers. We experimentally study 1580 DRAM chips (408$\times$ DDR3, 652$\times$ DDR4, and 520$\times$ LPDDR4) from 300 DRAM modules (60$\times$ DDR3, 110$\times$ DDR4, and 130$\times$ LPDDR4) and present our RowHammer characterization results for both aggregate RowHammer failure rates and the behavior of individual cells while sweeping the hammer count ($HC$) and stored data pattern. 
    \item Via our rigorous characterization studies, we definitively demonstrate that the RowHammer vulnerability significantly worsens (i.e., the number of hammers required to induce a RowHammer bit flip, {\hcfirst}, greatly reduces) in newer DRAM chips (e.g., {\hcfirst} reduces from $69.2k$ to $22.4k$ in DDR3, $17.5k$ to $10k$ in DDR4, and $16.8k$ to $4.8k$ in LPDDR4 chips across multiple technology node generations).
    \item We demonstrate, based on our rigorous evaluation of five state-of-the-art RowHammer mitigation mechanisms, that even though existing RowHammer mitigation mechanisms are reasonably effective at mitigating RowHammer in today's DRAM chips (e.g., 8\% average performance loss on our workloads when {\hcfirst} is 4.8k), they will cause significant overhead in future DRAM chips with even lower {\hcfirst} values (e.g., 80\% average performance loss with the most scalable mechanism when {\hcfirst} is 128).  
	\item We evaluate an ideal refresh-based mitigation mechanism that selectively refreshes a row only just before it is about to experience a RowHammer bit flip, and find that in chips with high vulnerability to RowHammer, there is still significant opportunity for developing a refresh-based RowHammer mitigation mechanism with low performance overhead that scales to low {\hcfirst} values. We conclude that it is critical to research more effective solutions to RowHammer, and we provide promising directions for future research. 
\end{itemize}


%% file: 2_background.tex
\section{DRAM Background}
\label{sec:background}

In this section, we describe the necessary background on DRAM organization and
operation to explain the RowHammer vulnerability and its implications
for real systems. For further detail, we refer the reader to prior studies on
DRAM~\cite{lee2015adaptive, khan2016parbor, seshadri2016simple,
hassan2016chargecache, hassan2017softmc, zhang2014half, lee2013tiered,
kim2012case, seshadri2013rowclone, chang2016understanding, lee2016reducing,
chang2017thesis, chang2016low, lee-sigmetrics2017, seshadri2017ambit,
lee2015decoupled, kim2016ramulator, kim2014flipping, lee2015simultaneous,
ghose2018your, chang2017understanding, ghose2019demystifying, seshadri2020indram, dennard1968field, kim2019d}.

\subsection{DRAM Organization}

\begin{figure}[h] \centering
    \includegraphics[width=0.85\linewidth]{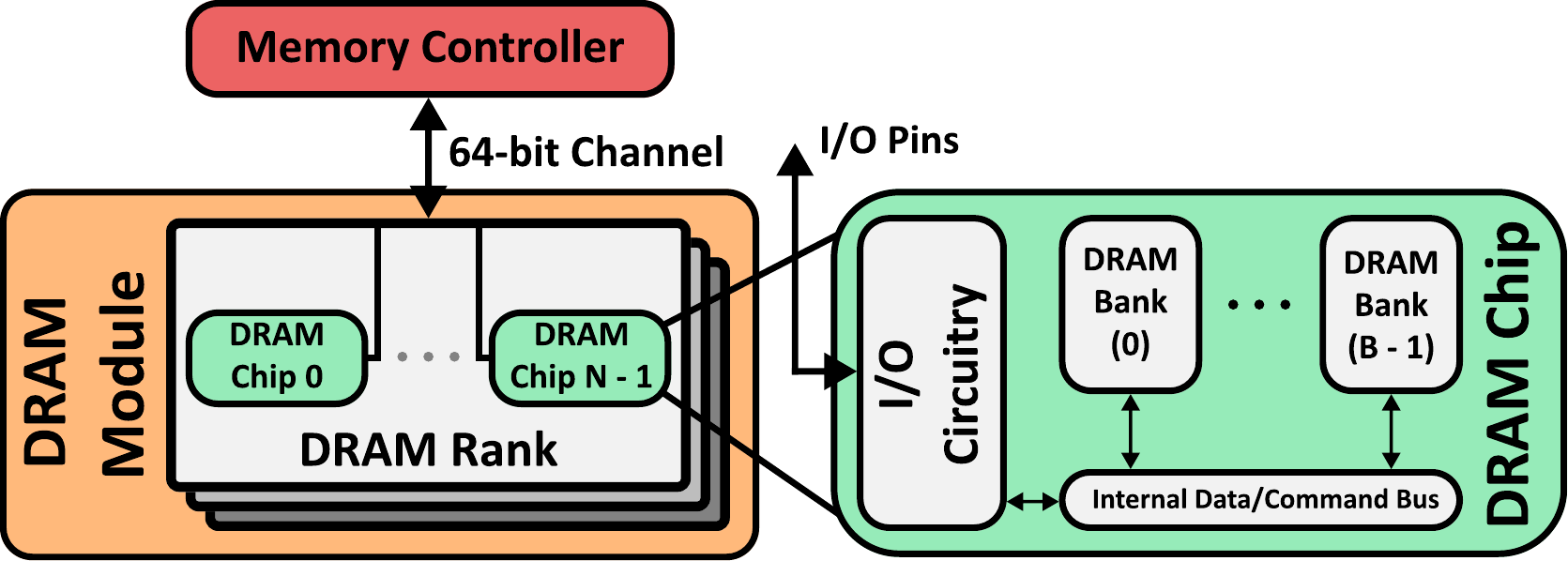}
    \caption{A typical DRAM-based system.} 
    \label{fig:dram_org}
\end{figure}

A typical computing system includes multiple DRAM \emph{channels},
where each channel has a separate I/O bus and operates independently of the
other channels in the system. As Figure~\ref{fig:dram_org}~(left)
illustrates, a memory controller can interface with multiple DRAM
\emph{ranks} by time-multiplexing the channel's I/O bus between the
ranks. Because the I/O bus is shared, the memory controller serializes
accesses to different ranks in the same channel.
A DRAM rank comprises multiple DRAM \emph{chips} that operate in
lockstep. The combined data pins from all chips form the DRAM data bus.

\textbf{DRAM Chip Organization.}
A modern DRAM chip contains billions of cells, each of which
stores a single bit of data. Within a chip, cells are organized
hierarchically to provide high density and performance.
As Figure~\ref{fig:dram_org}~(right) shows, a DRAM chip is composed of
multiple (e.g., 8--16~\cite{jedec2012, ddr4}) DRAM \emph{banks}. All
banks within a chip share the internal data and command bus.

\begin{figure}[h]
    \centering
    \begin{subfigure}[b]{.60\linewidth}
        \centering
        \includegraphics[width=\linewidth]{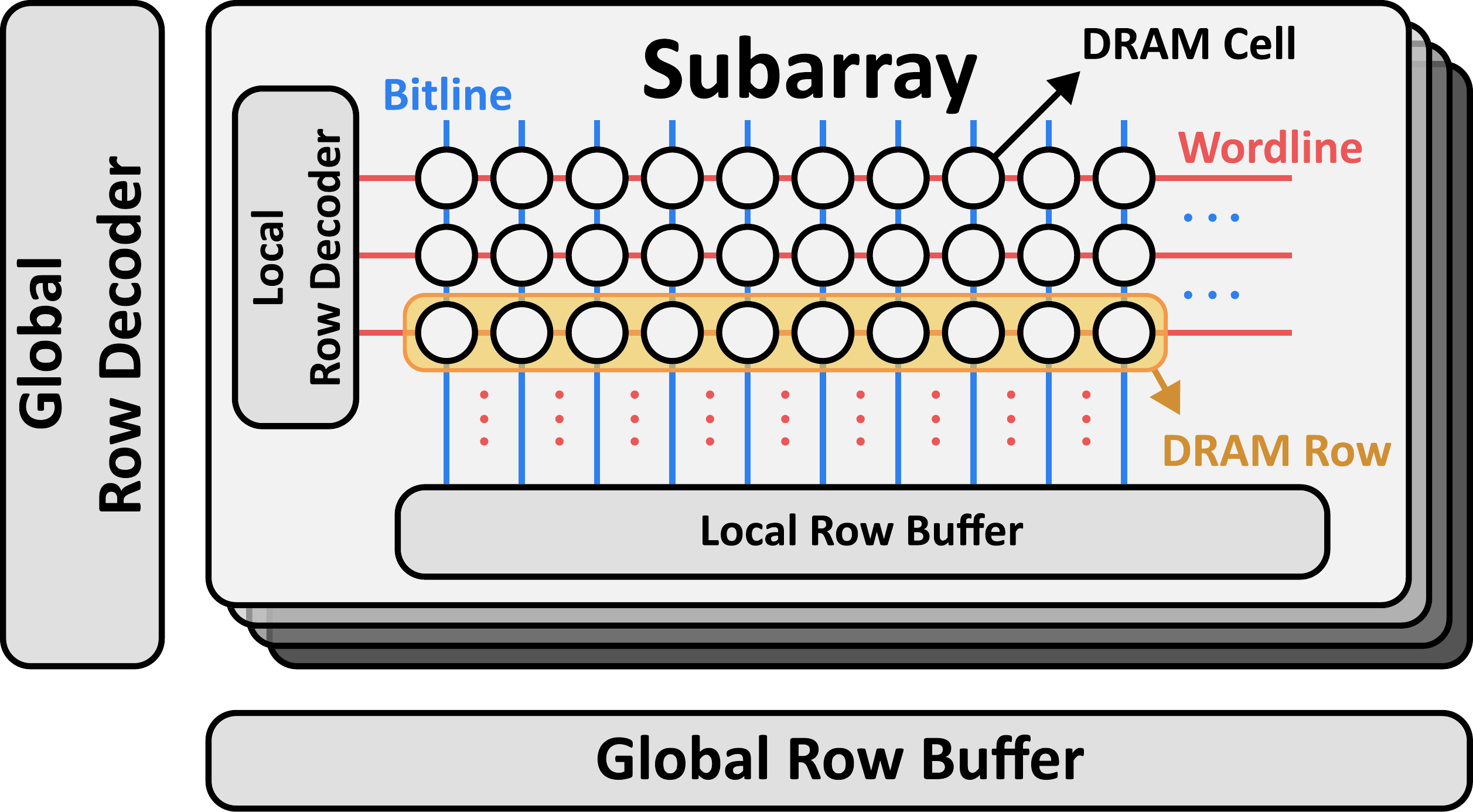}
        \label{subfig:dram_bank}
    \end{subfigure}
    \quad
    \begin{subfigure}[b]{.245\linewidth}
        \centering
        \includegraphics[width=\linewidth]{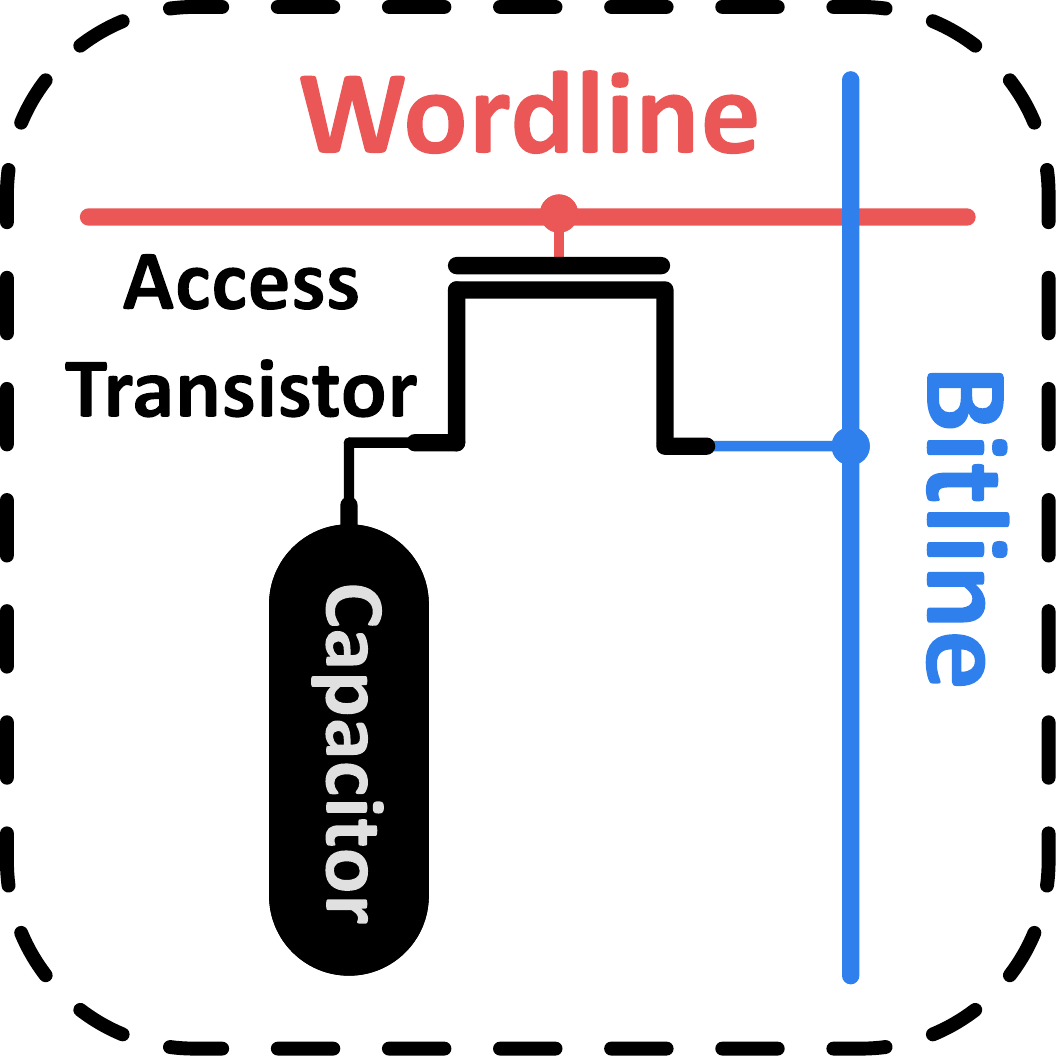}
        \label{subfig:dram_cell}
    \end{subfigure}
        \vspace{-4mm}
        \caption{DRAM bank and cell.} 
        \label{fig:dram_bank_and_cell}
\end{figure}

Figure~\ref{fig:dram_bank_and_cell}~(left) shows the internal organization of
a DRAM bank. A bank comprises many (e.g., 128) subarrays~\cite{kim2012case,
chang2016low}.  Each subarray contains a two-dimensional array of DRAM cells
arranged in \emph{rows} and \emph{columns}. When accessing DRAM, the memory
controller first provides the address of the row to be accessed. Then, the row
decoder, which is also hierarchically organized into global and local
components, opens the row by driving the corresponding \emph{wordline}. DRAM
cells that are connected to the same wordline are collectively referred to
as a DRAM \emph{row}.  To read and manipulate a cell's contents, a wire (i.e.,
\emph{bitline}) connects a column of cells to a \emph{sense amplifier}. The
collection of sense amplifiers in a subarray is referred to as the \emph{local
row buffer}.  The local row buffers of the subarrays in a bank are connected to
a per-bank \emph{global row buffer}, which interfaces with the internal command
and data bus of the DRAM chip.

As Figure~\ref{fig:dram_bank_and_cell}~(right) shows, a DRAM cell consists of
an \emph{access transistor} and a \emph{capacitor}. The wordline is connected
to the gate of the access transistor that, when enabled, connects the
cell capacitor to the bitline. A DRAM cell stores a single bit of data based
on the charge level of the cell capacitor (e.g., a charged
capacitor represents a logical value of ``1'' and a discharged capacitor
a value of ``0'', or vice versa). Unfortunately, charge leaks from the
storage capacitor over time due to various charge leakage paths in
the circuit components. To ensure that the cell does not leak enough charge
to cause a bit flip, a DRAM cell needs to be periodically
refreshed~\cite{liu2013experimental, raidr}.

\subsection{DRAM Operation}

A memory controller issues a sequence of DRAM commands to access data in a DRAM
chip. First, the memory controller issues an \emph{activate (ACT)} command to
\emph{open} a row that corresponds to the memory address to be accessed.
Opening (i.e., activating) a DRAM row causes the data in the target DRAM row to
be copied to its corresponding local row buffer. Second, the memory controller
issues either a \emph{READ} or a \emph{WRITE} command to the DRAM to read out
or update the target data within the row buffer, typically 32 or 64 bytes split
across all chips in the rank. The memory controller can issue multiple
consecutive \emph{READ} and \emph{WRITE} commands to an open row. While a row
is open, its cells remain connected to the sense amplifiers in the local row
buffer, so changes to the data stored in the row buffer propagate to the
DRAM cells. When accesses to the open row are complete, the memory controller
issues a \emph{precharge (PRE)} command to \emph{close} the open row and
prepare the bank to activate a different row.


\textbf{DRAM Refresh.} DRAM cell capacitors lose their charge over
time~\cite{liu2013experimental, raidr, saino2000impact}, potentially
resulting in bit flips. A cell's \emph{retention time} refers to the
duration for which its capacitor maintains the correct value. Cells
throughout a DRAM chip have different retention times, ranging from
milliseconds to hours~\cite{hamamoto1998retention, hassan2017softmc, kang2014co, khan2014efficacy,
kim2009new, lee2015adaptive, li2011dram, liu2013experimental, raidr,
patel2017reaper, qureshi2015avatar, venkatesan2006retention,
khan2017detecting, halderman2009lest}. To prevent data loss, the memory
controller issues regular \emph{refresh (REF)} commands that
ensure every DRAM cell is refreshed at fixed intervals (typically
every 32 or 64 ms according to DRAM specifications~\cite{2014lpddr4,
lpddr3, ddr4}).


\subsection{RowHammer: DRAM Disturbance Errors}

Modern DRAM devices suffer from \emph{disturbance errors} that
occur when a high rate of accesses to a single DRAM row unintentionally
flip the values of cells in nearby rows. This phenomenon is known as
\emph{RowHammer}~\cite{kim2014flipping}. It inherently stems from
electromagnetic interference between nearby cells. RowHammer is exacerbated by
reduction in process technology node size because adjacent DRAM cells
become both smaller and closer to each other.  Therefore, as DRAM manufacturers
continue to increase DRAM storage density, a chip's vulnerability to RowHammer
bit flips increases~\cite{kim2014flipping, mutlu2017rowhammer,
mutlu2019rowhammer}.

RowHammer exposes a system-level security vulnerability that has been studied
by many prior works both from the attack and defense perspectives. Prior
works demonstrate that RowHammer can be used to mount system-level attacks for
privilege escalation (e.g.,~\cite{cojocar2019exploiting,
gruss2018another, gruss2016rowhammer, lipp2018nethammer, qiao2016new,
razavi2016flip, seaborn2015exploiting, tatar2018defeating, van2016drammer,
xiao2016one, frigo2020trrespass, ji2019pinpoint}),
leaking confidential data (e.g.,~\cite{kwong2020rambleed}), and
denial of service (e.g.,~\cite{gruss2018another, lipp2018nethammer}). These
works effectively demonstrate that a system must provide protection against
RowHammer to ensure
robust (i.e., reliable and secure) execution.

Prior works propose defenses against RowHammer attacks both at the
hardware (e.g.,~\cite{kim2014flipping, lee2019twice, ryu2017overcoming,
son2017making, ghasempour2015armor, you2019mrloc, seyedzadeh2018cbt,
kang2020cat, kim2014architectural, gomez2016dram, bains2015rowref, bains14d, bains14c, greenfield14b, bains2016row, bains2015row, hassan2019crow, fisch2017dram}) and
software (e.g.,~\cite{rh-apple, aweke2016anvil, brasser2016can,
kim2014flipping, konoth2018zebram, li2019detecting, van2018guardion, rh-lenovo,
rh-hp, irazoqui2016mascat, bu2018srasa, rh-cisco, wu2019protecting,
bock2019rip, kim2019effective, wang2019reinforce, chakraborty2019deep,
wang2019detect}) levels. DRAM manufacturers themselves employ in-DRAM RowHammer
prevention mechanisms such as \emph{Target Row Refresh (TRR)}~\cite{ddr4},
which internally performs proprietary operations to reduce the vulnerability of
a DRAM chip against potential RowHammer attacks, although these solutions have
been recently shown to be vulnerable~\cite{frigo2020trrespass}. Memory
controller and system manufacturers have also included defenses such as
increasing the refresh rate~\cite{rh-apple, aweke2016anvil, rh-lenovo} and
Hardware RHP~\cite{intel2017cannon, omron2019ny, tq2020tq,
versalogic2019blackbird}.  For a detailed survey of the RowHammer problem, its
underlying causes, characteristics, exploits building on it, and mitigation
techniques, we refer the reader to~\cite{mutlu2019rowhammer}.

%% file: 3_motivation.tex
\section{Motivation and Goal} 
\label{sec:motivation}



Despite the considerable research effort expended towards understanding and
mitigating RowHammer, scientific literature still lacks rigorous
experimental data on how the RowHammer vulnerability is changing with the
advancement of DRAM designs and process technologies. In general, important
practical concerns are difficult to address with existing data in
literature. For example: 
\begin{itemize}
\item How vulnerable to RowHammer are future DRAM chips expected to be at the circuit level?
\item How well would RowHammer mitigation mechanisms prevent or mitigate RowHammer in future devices?
\item What types of RowHammer solutions would cope best with increased circuit-level vulnerability due to continued technology node scaling?
\end{itemize} 
While existing experimental characterization
studies~\cite{kim2014flipping, park2016statistical, park2016experiments} take
important steps towards building an overall understanding of the RowHammer
vulnerability, they are too scarce and collectively do not provide a holistic
view of RowHammer evolution into the modern day. To help overcome this lack of
understanding, we need a unifying study of the RowHammer vulnerability of a
broad range of DRAM chips spanning the time since the original RowHammer
paper was published in 2014~\cite{kim2014flipping}.


To this end, \textbf{our goal} in this paper is to evaluate and understand
how the RowHammer vulnerability of real DRAM chips at the circuit level
changes across different chip types, manufacturers, and process technology
node generations. Doing so enables us to predict how the RowHammer
vulnerability in DRAM chips will scale as the industry continues to increase
storage density and reduce technology node size for future chip designs. To
achieve this goal, we perform a rigorous experimental characterization study of
DRAM chips from three different DRAM types (i.e., DDR3, DDR4, and LPDDR4),
three major DRAM manufacturers, and at least two different process technology
nodes from each DRAM type. We show how different chips from different DRAM
types and technology nodes (abbreviated as ``type-node'' configurations)
have varying levels of vulnerability to RowHammer. We compare the chips'
vulnerabilities against each other and
project how they will likely scale when reducing the technology node size even
further (Section~\ref{sec:characterization}). Finally, we study how effective
existing RowHammer mitigation mechanisms will be, based on our observed
and projected experimental data on the RowHammer vulnerability
(Section~\ref{sec:implications}).

%% file: 4_methodology.tex
\section{Experimental Methodology}
\label{sec:methodology}

We describe our methodology for characterizing DRAM chips for RowHammer. 

\subsection{Testing Infrastructure}
\label{subsec:methodology:infrastructure}

In order to characterize the effects of RowHammer across a broad range of
modern DRAM chips, we experimentally study DDR3, DDR4, and LPDDR4 DRAM chips
across a wide range of testing conditions. To achieve this, we use two
different testing infrastructures: (1) the SoftMC
framework~\cite{softmc-safarigithub, hassan2017softmc} capable of testing DDR3
and DDR4 DRAM modules in a temperature-controlled chamber and (2) an in-house
temperature-controlled testing chamber capable of testing LPDDR4 DRAM chips.

\begin{figure}[h] \centering
    \includegraphics[width=0.85\linewidth]{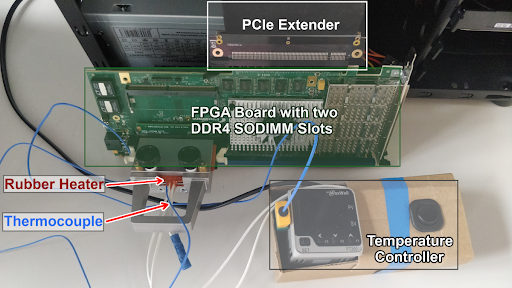}
    \caption{Our SoftMC infrastructure~\cite{softmc-safarigithub, hassan2017softmc} for testing DDR4 DRAM chips.}
    \label{fig:softmc_ddr4}
\end{figure}

\textbf{SoftMC.} Figure~\ref{fig:softmc_ddr4} shows our SoftMC setup for
testing DDR4 chips. In this setup, we use an FPGA board with a Xilinx Virtex
UltraScale 95 FPGA~\cite{xilinx_virtex_ultrascale}, two DDR4 SODIMM slots, and
a PCIe interface. To open up space around the DDR4 chips for temperature
control, we use a vertical DDR4 SODIMM riser board to plug a DDR4 module into the
FPGA board.  We heat the DDR4 chips to a target temperature using silicone
rubber heaters pressed to both sides of the DDR4 module. We control the temperature
using a thermocouple, which we place between the rubber heaters and the DDR4
chips, and a temperature controller. To enable fast data transfer between the
FPGA and a host machine, we connect the FPGA to the host machine using PCIe via a 30~cm
PCIe extender. We use the host machine to program the SoftMC hardware and
collect the test results. Our SoftMC setup for testing DDR3 chips is similar
but uses a Xilinx ML605 FPGA board~\cite{xilinx_ml605}. Both infrastructures
provide fine-grained control over the types and timings of DRAM commands sent to
the chips under test and provide precise temperature control at typical
operating conditions. 

\textbf{LPDDR4 Infrastructure.} Our LPDDR4 DRAM testing infrastructure uses
industry-developed in-house testing hardware for package-on-package LPDDR4
chips. The LPDDR4 testing infrastructure is further equipped with cooling
and heating capabilities that also provide us with precise temperature
control at typical operating conditions.  

\subsection{Characterized DRAM Chips}
\label{subsec:methodology:devices}

Table~\ref{tab:devices} summarizes the 1580 DRAM chips that we test using both
infrastructures. We have chips from all of the three major DRAM manufacturers
spanning DDR3 (listed in Appendix Table~\ref{table:ddr3_table}), DDR4 (listed
in Appendix Table~\ref{table:ddr4_table}), and two known technology nodes of
LPDDR4. We refer to the DRAM type (e.g., LPDDR4) and technology node of a DRAM
chip as a \emph{DRAM type-node configuration} (e.g., LPDDR4-1x). For DRAM chips
whose technology node we do not exactly know, we identify their node as
\emph{old} or \emph{new}.
\begin{table}[h!]
\footnotesize 
\begin{center}
\caption{Summary of DRAM chips tested.} 
\begin{tabular}{ lrrrr }
\toprule
\multicolumn{1}{c}{\textbf{DRAM}} & \multicolumn{4}{c}{\textbf{Number of Chips (Modules) Tested}} \\
\multicolumn{1}{c}{\textbf{type-node}} & \textbf{\textit{Mfr. A}} & \textbf{\textit{Mfr. B}} & \textbf{\textit{Mfr. C}} & \textbf{\textit{Total}} \\
\toprule
DDR3-old & 56 (10) & 88 (11) & 28 (7) & \textbf{172 (28)} \\
DDR3-new & 80 (10) & 52 (9) & 104 (13) & \textbf{236 (32)} \\
\Xhline{0.5\arrayrulewidth}
DDR4-old & 112 (16) & 24 (3) & 128 (18) & \textbf{264 (37)} \\
DDR4-new & 264 (43) & 16 (2) & 108 (28) & \textbf{388 (73)} \\
\Xhline{0.5\arrayrulewidth}
LPDDR4-1x & 12 (3) & 180 (45) & N/A & \textbf{192 (48)} \\
LPDDR4-1y & 184 (46) & N/A & 144 (36) & \textbf{328 (82)} \\ 
\toprule
\end{tabular}
\label{tab:devices}
\vspace{-3mm} 
\end{center}
\end{table}


\textbf{DDR3 and DDR4.} Among our tested DDR3 modules, we identify two distinct
batches of chips based on their manufacturing date, datasheet publication date,
purchase date, and RowHammer characteristics. We categorize
DDR3 devices with a manufacturing date earlier than 2014 as DDR3-old chips, and
devices with a manufacturing date including and after 2014 as DDR3-new chips.
Using the same set of properties, we identify two distinct batches of devices
among the DDR4 devices. We categorize DDR4 devices with a manufacturing date
before 2018 or a datasheet publication date of 2015 as DDR4-old chips and
devices with a manufacturing date including and after 2018 or a datasheet
publication date of 2016 or 2017 as DDR4-new chips. Based on our observations
on RowHammer characteristics from these chips, we expect that DDR3-old/DDR4-old
chips are manufactured at an older date with an older process technology
compared to DDR3-new/DDR4-new chips, respectively. This enables us to directly
study the effects of shrinking process technology node sizes in DDR3 and DDR4
DRAM chips. 

\textbf{LPDDR4.} For our LPDDR4 chips, we have two known distinct
generations manufactured with different technology node sizes, 1x-nm and 1y-nm,
where 1y-nm is smaller than 1x-nm.  Unfortunately, we are missing data from
some generations of DRAM from specific manufacturers (i.e., LPDDR4-1x from
manufacturer C and LPDDR4-1y from manufacturer B) since we did not have access
to chips of these manufacturer-technology node combinations due to
confidentiality issues. Note that while we know the external technology node
values for the chips we characterize (e.g., 1x-nm, 1y-nm), these values are
\emph{not} standardized across different DRAM manufacturers and the actual
values are confidential.  This means that a 1x chip from one manufacturer is
not necessarily manufactured with the same process technology node as a 1x chip
from another manufacturer. However, since we do know relative process node
sizes of chips from the \emph{same} manufacturer, we can directly observe how
technology node size affects RowHammer on LPDDR4 DRAM chips. 

\subsection{Effectively Characterizing RowHammer}
\label{subsec:methodology:characterizing_rowhammer}

In order to characterize RowHammer effects on our DRAM chips at the
circuit-level, we want to test our chips at the worst-case RowHammer
conditions. We identify two conditions that our tests must satisfy to
effectively characterize RowHammer at the circuit level: our testing
routines must both: 1) run without interference (e.g., without DRAM refresh or
RowHammer mitigation mechanisms) and 2) systematically test each DRAM row's
vulnerability to RowHammer by issuing the \emph{worst-case sequence of DRAM
accesses} for that particular row. 

\noindent\textbf{Disabling Sources of Interference.} To directly observe
RowHammer effects at the circuit level, we want to minimize the external
factors that may limit 1) the effectiveness of our tests or 2) our ability
to effectively characterize/observe circuit-level effects of RowHammer on our
DRAM chips.  First, we want to ensure that we have control over how our
RowHammer tests behave without disturbing the desired access pattern in any way. Therefore, during
the core loop of each RowHammer test (i.e., when activations are issued at a
high rate to induce RowHammer bit flips), we disable all DRAM
self-regulation events such as refresh and calibration, using control registers
in the memory controller. This guarantees consistent testing without
confounding factors due to intermittent events (e.g., to avoid the
possibility that a victim row is refreshed during a RowHammer test routine such
that we observe fewer RowHammer bit flips).  Second, we want to directly
observe the circuit-level bit flips such that we can make conclusions about
DRAM's vulnerability to RowHammer at the circuit technology level rather than
the system level. To this end, to the best of our knowledge, we disable all
DRAM-level (e.g., TRR~\cite{2014lpddr4, ddr4, frigo2020trrespass}) and
system-level RowHammer mitigation mechanisms (e.g.,
pTRR~\cite{aichinger2015ddr}) along with all forms of rank-level
error-correction codes (ECC), which could obscure RowHammer bit flips.
Unfortunately, all of our LPDDR4-1x and LPDDR4-1y chips use on-die
ECC~\cite{micron2017whitepaper, kwak2017a, kang2014co, patel2019understanding,
kwon2017an} (i.e., an error correcting mechanism that corrects single-bit
failures entirely within the DRAM chip~\cite{patel2019understanding}), which we
cannot disable. Third, we ensure that the core loop of our RowHammer test
runs for less than 32 ms (i.e., the lowest refresh interval specified by
manufacturers to prevent DRAM data retention failures across our tested
chips~\cite{patel2017reaper, liu2013experimental, khan2014efficacy, jedec2012,
ddr4, 2014lpddr4}) so that we do not conflate retention failures with RowHammer
bit flips. 

\noindent\textbf{Worst-case RowHammer Access Sequence.} We leverage
\emph{three} key observations from prior work~\cite{kim2014flipping,
aweke2016anvil, gruss2018another, xiao2016one, cojocar2020we} in order to craft
a worst-case RowHammer test pattern. First, a repeatedly accessed row (i.e.,
\emph{aggressor row}) has the greatest impact on its immediate
physically-adjacent rows (i.e., repeatedly accessing physical row $N$ will
cause the highest number of RowHammer bit flips in physical rows $N+1$ and
$N-1$). Second, a \emph{double-sided hammer} targeting physical victim row $N$
(i.e., repeatedly accessing physical rows $N-1$ and $N+1$) causes the
\emph{highest} number of RowHammer bit flips in row $N$ compared to any other
access pattern. Third, increasing the rate of DRAM activations (i.e., issuing
the same number of activations within shorter time periods) results in an
increasing number of RowHammer bit flips. This rate of activations is limited
by the DRAM timing parameter $t_{RC}$ (i.e., the time between two successive
activations) which depends on the DRAM clock frequency and the DRAM type: DDR3
(52.5ns)~\cite{jedec2012}, DDR4 (50ns)~\cite{ddr4}, LPDDR4
(60ns)~\cite{2014lpddr4}. Using these observations, we test each row's
worst-case vulnerability to RowHammer by repeatedly accessing the two directly
physically-adjacent rows as fast as possible.

To enable the quick identification of physical rows $N-1$ and $N+1$ for a
given row $N$, we reverse-engineer the \emph{undocumented} and
\emph{confidential} logical-to-physical DRAM-internal row address
remapping. To do this, we exploit RowHammer's key observation that
repeatedly accessing an arbitrary row causes the two directly
physically-adjacent rows to contain the \emph{highest} number of RowHammer bit
flips~\cite{kim2014flipping}. By repeating this analysis across rows throughout
the DRAM chip, we can deduce the address mappings for each type of chip that we
test. We can then use this mapping information to quickly test RowHammer
effects at worst-case conditions. We note that for our LPDDR4-1x chips from
Manufacturer B, when we repeatedly access a single row within two consecutive
rows such that the first row is an even row (e.g., rows 2 and 3) in the logical
row address space as seen by the memory controller, we observe 1) no RowHammer
bit flips in either of the two consecutive rows and 2) a near equivalent number
of RowHammer bit flips in each of the four immediately adjacent rows: the two
previous consecutive rows (e.g., rows 0 and 1) and the two subsequent
consecutive rows (e.g., rows 4 and 5).  This indicates a row address remapping
that is internal to the DRAM chip such that every pair of consecutive rows
share the same internal wordline.  To account for this DRAM-internal row
address remapping, we test each row $N$ in LPDDR4-1x chips from manufacturer B
by repeatedly accessing physical rows $N-2$ and $N+2$.

\noindent\textbf{Additional Testing Parameters.} To investigate RowHammer
characteristics, we explore two testing parameters at a stable ambient
temperature of $50^{\circ}C$:  
\begin{enumerate}
    \item \textbf{Hammer count ($HC$).} We test the effects of changing the number of times we access (i.e., activate) a victim row's physically-adjacent rows (i.e., aggressor rows). We count each pair of activations to the two neighboring rows as one \emph{hammer} (e.g., one activation each to rows $N-1$ and $N+1$ counts as one hammer). We sweep the hammer count from 2k to 150k (i.e., 4k to 300k activations) across our chips so that the hammer test runs for less than 32ms. 
    \item \textbf{Data pattern ($DP$).} We test several commonly-used DRAM data patterns where every byte is written with the same data: Solid0 (SO0: 0x00), Solid1 (SO1: 0xFF), Colstripe0 (CO0: 0x55), Colstripe1 (CO1: 0xAA)~\cite{liu2013experimental, patel2017reaper, khan2014efficacy}. In addition, we test data patterns where each byte in every other row, including the row being hammered, is written with the same data, Checkered0 (CH0: 0x55) or Rowstripe0 (RS0: 0x00), and all other rows are written with the inverse data, Checkered1 (CH1: 0xAA) or Rowstripe1 (RS1: 0xFF), respectively.
\end{enumerate} 

\noindent\textbf{RowHammer Testing Routine.}
Algorithm~\ref{alg:rowhammer_char} presents the general testing methodology we
use to characterize RowHammer on DRAM chips. For different data patterns
($DP$) (line~2) and hammer counts ($HC$) 
(line~8), the test individually targets each row in DRAM (line~4) as
a victim row (line~5).  For each victim row, we identify the two
physically-adjacent rows ($aggressor\_row1$ and $aggressor\_row2$) as aggressor
rows (lines~6 and 7).  Before beginning the core loop of our RowHammer test
(Lines 11-13), two things happen: 1) the memory controller disables DRAM
refresh (line~9) to ensure no interruptions in the core loop of our test due to
refresh operations, and 2) we refresh the victim row (line~10) so that we
begin inducing RowHammer bit flips on a fully-charged row, which ensures that
bit flips we observe are not due to retention time violations. The core loop of our
RowHammer test (Lines 11-13) induces RowHammer bit flips in the victim row
    \begin{algorithm}[tbh]\footnotesize
        \setstretch{0.6} 
        \SetAlgoNlRelativeSize{0.7}
        \SetAlgoNoLine
        \DontPrintSemicolon
        \SetAlCapHSkip{0pt}
        \caption{DRAM RowHammer Characterization}
        \label{alg:rowhammer_char}
        \textbf{DRAM\_RowHammer\_Characterization():} \par
		\quad\quad \textbf{foreach} $DP$ in [Data Patterns]: \par
        \quad\quad\quad\quad write $DP$ into all cells in $DRAM$ \par
		\quad\quad\quad\quad\textbf{foreach} $row$ in $DRAM$: \par
		\quad\quad\quad\quad\quad\quad set $victim\_row$ to $row$ \par
		\quad\quad\quad\quad\quad\quad set $aggressor\_row1$ to $victim\_row - 1$ \par
		\quad\quad\quad\quad\quad\quad set $aggressor\_row2$ to $victim\_row + 1$ \par
		\quad\quad\quad\quad\quad\quad \textbf{foreach} $HC$ in [$HC$ sweep]: \par
		\quad\quad\quad\quad\quad\quad\quad\quad Disable DRAM refresh \par
		\quad\quad\quad\quad\quad\quad\quad\quad Refresh $victim\_row$ \par
		\quad\quad\quad\quad\quad\quad\quad\quad \textbf{for} $n = 1 \rightarrow HC$: \textcolor{gray}{// core test loop} \par
		\quad\quad\quad\quad\quad\quad\quad\quad\quad\quad activate $aggressor\_row1$ \par
		\quad\quad\quad\quad\quad\quad\quad\quad\quad\quad activate $aggressor\_row2$ \par
		\quad\quad\quad\quad\quad\quad\quad\quad Enable DRAM refresh \par
        \quad\quad\quad\quad\quad\quad\quad\quad Record RowHammer bit flips to storage \par
		\quad\quad\quad\quad\quad\quad\quad\quad Restore bit flips to original values \par 
    \end{algorithm}
by first activating $aggressor\_row1$ then $aggressor\_row2$, $HC$ times. After
the core loop of our RowHammer test, we re-enable DRAM refresh (line~14) to
prevent retention failures and record the observed bit flips to secondary
storage (line~15) for analysis (presented in
Section~\ref{sec:characterization}). Finally, we prepare to test the next $HC$
value in the sweep by restoring the observed bit flips to their original values
(Line 16) depending on the data pattern ($DP$) being tested. 

\textbf{Fairly Comparing Data Across Infrastructures.} Our
carefully-crafted RowHammer test routine allows us to compare our test results
between the two different testing infrastructures.
This is because, as we described earlier, we 1) reverse engineer the row address mappings of each DRAM
configuration such that we effectively test double-sided RowHammer on every
single row, 2) issue activations as fast as possible for each chip, such that
the activation rates are similar across infrastructures, and 3) disable all
sources of interference in our RowHammer tests. 


%% file: 5_characterization.tex
\section{RowHammer Characterization}
\label{sec:characterization}

\newcounter{obscount}

In this section, we present our comprehensive characterization of RowHammer on
the 1580 DRAM chips we test. 

\subsection{RowHammer Vulnerability}
\label{subsec:vulnerability}

We first examine which of the chips that we test are susceptible to RowHammer.
Across all of our chips, we sweep the hammer count ($HC$) between 2K and
150K (i.e., 4k and 300k activates for our double-sided RowHammer test) and
observe whether we can induce any RowHammer bit flips at all in each chip.  We
find that we can induce RowHammer bit flips in all chips except many
DDR3 chips. Table~\ref{tab:ddr3_failing_devices} shows the fraction of DDR3
chips in which we \emph{can} induce RowHammer bit flips (i.e., \emph{RowHammerable}
chips). 

\begin{table}[h!]
\footnotesize
\begin{center}
\caption{Fraction of DDR3 DRAM chips vulnerable to RowHammer when $\bm{HC}\, \mathbf{<150k}$.} 
\begin{tabular}{ lrrr}
\toprule
\multicolumn{1}{c}{\textbf{DRAM}} & \multicolumn{3}{c}{\textbf{RowHammerable chips}} \\
\multicolumn{1}{c}{\textbf{type-node}} & \textbf{\textit{Mfr. A}} & \textbf{\textit{Mfr. B}} & \textbf{\textit{Mfr. C}} \\
\toprule
DDR3-old & 24/88 & 0/88 & 0/28 \\
DDR3-new & 8/72 & 44/52 & 96/104 \\
\toprule
\end{tabular}
\label{tab:ddr3_failing_devices}
\vspace{-3mm} 
\end{center}
\end{table}

\stepcounter{obscount} \textbf{Observation \arabic{obscount}.} \emph{Newer
DRAM chips appear to be more vulnerable to RowHammer based on the increasing
fraction of RowHammerable chips from DDR3-old to DDR3-new DRAM chips of
manufacturers B and C.} 

We find that the fraction of manufacturer A's chips that are RowHammerable
decreases from DDR3-old to DDR3-new chips, but we also note that the number of
RowHammer bit flips that we observe across each of manufacturer A's chips is
very low ($<20$ on average across RowHammerable chips) compared to the number
of bit flips found in manufacturer B and C's DDR3-new chips ($87k$ on average
across RowHammerable chips) when $HC=150K$. Since DDR3-old chips of all
manufacturers and DDR3-new chips of manufacturer A have very few to no bit
flips, we refrain from analyzing and presenting their characteristics in many
plots in Section~\ref{sec:characterization}. 

%
%


\subsection{Data Pattern Dependence}
\label{subsec:dpd}

To study data pattern effects on observable RowHammer bit flips, we
test our chips using Algorithm~\ref{alg:rowhammer_char} with $hammer\_count \
(HC)=150k$ at $50^{\circ}C$, sweeping the 1) $victim\_row$ and 2)
$data\_pattern$ (as described in
Section~\ref{subsec:methodology:characterizing_rowhammer}).\footnote{Note
that for a given data pattern ($DP$), the same data is always written to
$victim\_row$. For example, when testing Rowstripe0, every byte in
$victim\_row$ is always written with 0x00 and every byte in the two
physically-adjacent rows are written with 0xFF.} 

\begin{figure}[h] \centering
    \includegraphics[width=0.93\linewidth]{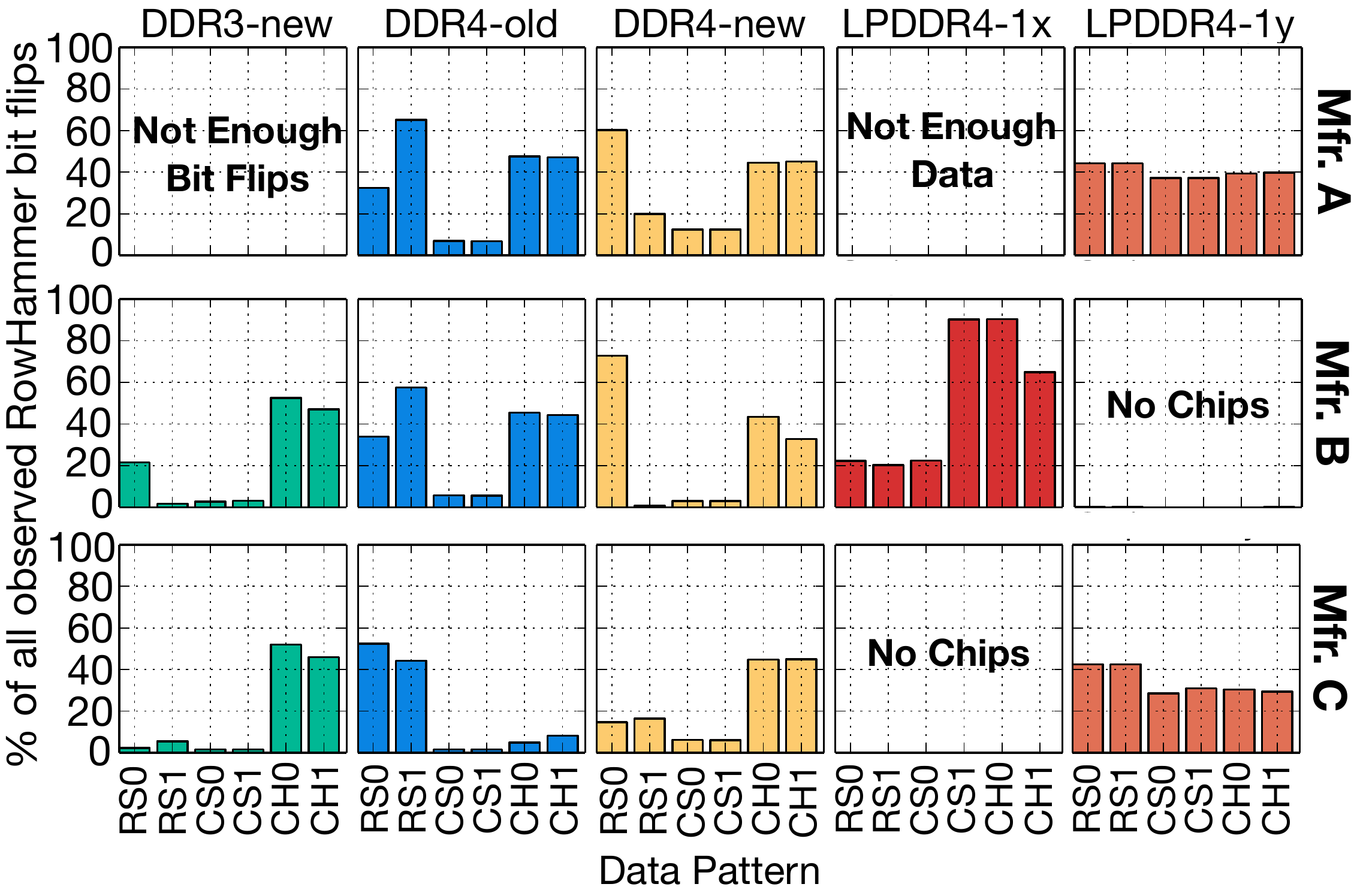}
	\caption{RowHammer bit flip coverage of different data patterns (described in Section~\ref{subsec:methodology:characterizing_rowhammer}) for a single representative DRAM chip of each type-node configuration.} 
    \label{fig:char:dpd_distribution}
\end{figure}

We first examine the set of all RowHammer bit flips that we observe when
testing with different data patterns for a given $HC$.
For each data pattern, we run our RowHammer test routine ten times. We then
aggregate all unique RowHammer bit flips per data pattern.
We combine all unique RowHammer bit flips found by all data patterns and
iterations into a full set of observable bit flips. Using the combined data,
we calculate the fraction of the full set of observable bit flips that each
data pattern identifies (i.e., the data pattern's \emph{coverage}).
Figure~\ref{fig:char:dpd_distribution} plots the coverage (y-axis) per individual data pattern (shared x-axis) for a
single representative DRAM chip from each DRAM type-node
configuration that we test. Each row of subplots shows the coverages for chips of the same
manufacturer (indicated on the right y-axis), and the columns show the
coverages for chips of the same DRAM type-node configuration (e.g., DDR3-new).

\stepcounter{obscount} \textbf{Observation \arabic{obscount}.}
\emph{Testing with different data patterns is essential for
comprehensively identifying RowHammer bit flips because no individual data
pattern achieves full coverage alone.} 

\stepcounter{obscount} \textbf{Observation \arabic{obscount}.}
\emph{The worst-case data pattern (shown in
Table~\ref{tab:dpd}) is consistent across chips of the same manufacturer and
DRAM type-node configuration.}\footnote{We do not consider the true/anti cell
pattern of a chip~\cite{liu2013experimental, kim2014flipping,
frigo2020trrespass} and agnostically program the data pattern accordingly into
the DRAM array. More RowHammer bit flips can be induced by considering the
true/anti-cell pattern of each chip and devising corresponding data patterns to
exploit this knowledge~\cite{frigo2020trrespass}.} 

\begin{table}[h]
\footnotesize
\begin{center}
\caption{Worst-case data pattern for each DRAM type-node configuration at $\mathbf{50^{\circ}C}$ split into different manufacturers.}
\begin{tabular}{ lrrr}
\toprule
\multicolumn{1}{c}{\textbf{DRAM}} & \multicolumn{3}{c}{\textbf{Worst Case Data Pattern at $\mathbf{50^{\circ}C}$}} \\
\multicolumn{1}{c}{\textbf{type-node}} & \textbf{\textit{Mfr. A}} & \textbf{\textit{Mfr. B}} & \textbf{\textit{Mfr. C}} \\
\toprule
DDR3-new & N/A & Checkered0 & Checkered0 \\
\Xhline{0.2\arrayrulewidth}
DDR4-old & RowStripe1 & RowStripe1 & RowStripe0 \\
DDR4-new & RowStripe0 & RowStripe0 & Checkered1 \\
\Xhline{0.2\arrayrulewidth}
\ap{LPDDR4-1x} & Checkered1 & Checkered0 & N/A \\
\ap{LPDDR4-1y} & RowStripe1 & N/A & RowStripe1 \\
\toprule
\end{tabular}
\label{tab:dpd}
\vspace{-3mm} 
\end{center}
\end{table}

We believe that different data patterns induce the most RowHammer bit flips in
different chips because DRAM manufacturers apply a variety of proprietary
techniques for DRAM cell layouts to maximize the cell density for different
DRAM type-node configurations. For the remainder of this
paper, we characterize each chip using \emph{only} its worst-case data
pattern.\footnote{We use the worst-case data pattern to 1) minimize the
extensive testing time, 2) induce many RowHammer bit flips, and 3) experiment
at worst-case conditions. A diligent attacker would also try to find the
worst-case data pattern to maximize the probability of a successful
RowHammer attack.}

\subsection{Hammer Count ($HC$) Effects} 
\begin{figure}[htbp] \centering
    \includegraphics[width=0.95\linewidth]{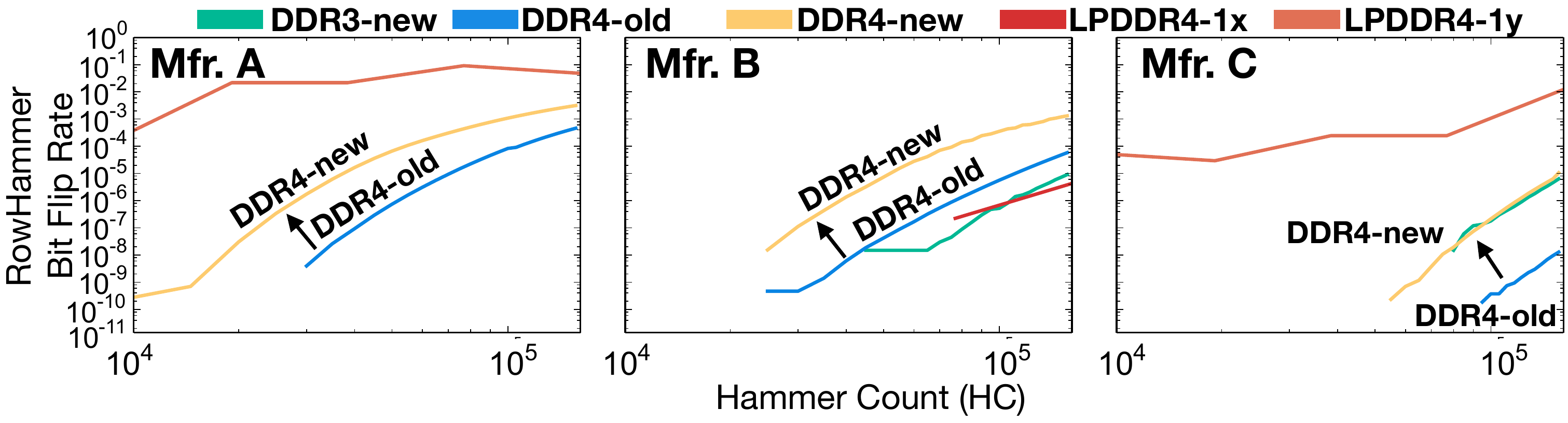}
    \caption{Hammer count ($HC$) vs. RowHammer bit flip rate across DRAM type-node configurations.} 
    \label{fig:char:refresh_sweep}
\end{figure}

We next study the effects of increasing the hammer count on the number of
observed RowHammer bit flips across our chips.
Figure~\ref{fig:char:refresh_sweep} plots the effects of increasing the number
of hammers on the RowHammer bit flip rate\footnote{We define the RowHammer bit
flip rate as the number of observed RowHammer bit flips to the total number of
bits in the tested DRAM rows.} for our tested DRAM chips of various DRAM
type-node configurations across the three major DRAM manufacturers. For all
chips, we hammer each row, sweeping $HC$ between 10,000 and 150,000.
For each $HC$ value, we plot the average rate of observed
RowHammer bit flips across all chips of a DRAM type-node configuration.

\stepcounter{obscount} \textbf{Observation \arabic{obscount}.} \emph{The log of
the number of RowHammer bit flips has a linear relationship with the log of
$HC$.}\footnote{Our observation is consistent with prior
work~\cite{park2016statistical}.}

We observe this relationship between $HC$ and RowHammer bit flip rate because
more accesses to a single row results in more cell-to-cell interference,
and therefore more charge is lost in victim cells of nearby rows. 

We examine the effects of DRAM technology node on the RowHammer bit flip rate
in Figure~\ref{fig:char:refresh_sweep}.  We observe that the bit flip rate
curve shifts \emph{upward} and \emph{leftward} when going from DDR4-old to DDR4-new chips,
indicating respectively, 1) a higher rate of bit flips for the same $HC$ value
and 2) occurrence of bit flips at lower $HC$ values, as technology node size
reduces from DDR4-old to DDR4-new.  

\stepcounter{obscount} \textbf{Observation \arabic{obscount}.} \emph{Newer
DDR4 DRAM technology nodes show a clear trend of increasing RowHammer bit flip
rates: the \emph{same} $HC$ value causes an increased average RowHammer bit
flip rate from DDR4-old to DDR4-new DRAM chips of all DRAM manufacturers.}

We believe that due to increased density of DRAM chips from older to newer
technology node generations, cell-to-cell interference increases and
results in DRAM chips that are more vulnerable to RowHammer bit flips.


\subsection{RowHammer Spatial Effects}
\label{subsec:spatial_effects} 

\begin{figure}[H] \centering
    \includegraphics[width=0.91\linewidth]{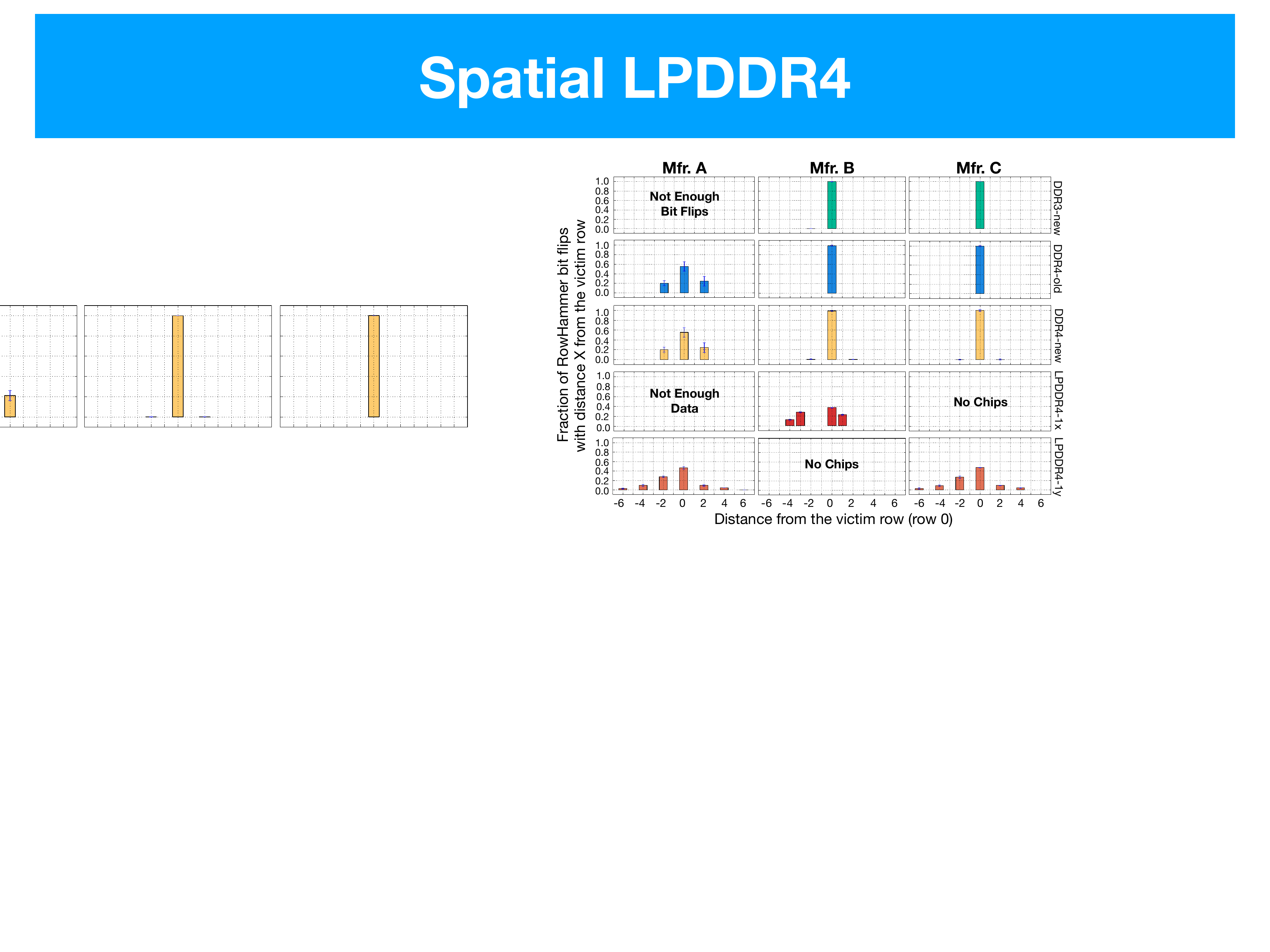}
    \caption{Distribution of RowHammer bit flips across row offsets from the victim row.} 
    \label{fig:char:distances}
\end{figure}

We next experimentally study the spatial distribution of RowHammer bit flips
across our tested chips. In order to normalize the RowHammer effects that we
observe across our tested chips, we first take each DRAM chip and use a hammer
count specific to that chip to result in a RowHammer bit flip rate of
$10^{-6}$.\footnote{We choose a RowHammer bit flip rate of $10^{-6}$ since we
are able to observe this bit flip rate in most chips that we characterize with
$HC<150k$.} For each chip, we analyze the spatial distribution of bit flips throughout the chip.
Figure~\ref{fig:char:distances} plots the fraction of RowHammer bit flips
that occur in a given row offset from the $victim\_row$ out of all observed
RowHammer bit flips. Each column of subplots shows the distributions for chips
of different manufacturers and each row of subplots shows the distribution for
a different DRAM type-node configuration. The error bars show the
standard deviation of the distribution across our tested chips. Note that the 
repeatedly-accessed rows (i.e., \emph{aggressor rows}) are
at $x=1$ and $x=-1$ for all plots except in LPDDR4-1x chips from
manufacturer B, where they are at $x=-2$ and $x=2$ (due to the internal address
remapping that occurs in these chips as we describe in
Section~\ref{subsec:methodology:characterizing_rowhammer}).  Because an access
to a row essentially refreshes the data in the row, repeatedly accessing
aggressor rows during the core loop of the RowHammer test prevents any bit
flips from happening in the aggressor rows. Therefore, there are no RowHammer
bit flips in the aggressor rows across each DRAM chip in our plots (i.e., $y=0$
for $x=[-2,-1,2,3]$ for LPDDR4-1x chips from manufacturer B and for $x=1$ and
$x=-1$ for all other chips). 

We make three observations from Figure~\ref{fig:char:distances}. First, we
observe a general trend across DRAM type-node configurations of a given DRAM
manufacturer where newer DRAM technology nodes have an increasing number of
rows that are susceptible to RowHammer bit flips that are \emph{farther} from
the victim row. For example, in LPDDR4-1y chips, we observe RowHammer bit flips
in as far as 6 rows from the victim row (i.e., $x=-6$), whereas in DDR3 and
DDR4 chips, RowHammer bit flips only occur in as far as 2 rows from the victim
row (i.e., $x=-2$). We believe that this effect could be due to 1) an
increase in DRAM cell density, which leads to cell-to-cell interference
extending farther than a single row, with RowHammer bit flips occurring in rows
increasingly farther away from the aggressor rows (e.g., 5 rows away) for
higher-density chips, and 2) more shared structures internal to
the DRAM chip, which causes farther (and multiple) rows to be affected by
circuit-level interference. 

\stepcounter{obscount} \textbf{Observation \arabic{obscount}.} \emph{For a
given DRAM manufacturer, chips of newer DRAM technology nodes can exhibit
RowHammer bit flips 1) in more rows and 2) farther away from the victim row.} 

Second, we observe that rows containing RowHammer bit flips that are
farther from the victim row have fewer RowHammer bit flips than rows closer to
the victim row. Non-victim rows adjacent to the aggressor rows ($x=2$ and
$x=-2$) contain RowHammer bit flips, and these bit flips demonstrate the
effectiveness of a single-sided RowHammer attack as only one of their adjacent
rows are repeatedly accessed. As discussed earlier
(Section~\ref{subsec:methodology:characterizing_rowhammer}), the single-sided
RowHammer attack is not as effective as the double-sided RowHammer attack, and
therefore we find fewer bit flips in these rows. In rows farther 
away from the victim row, we attribute the diminishing number of RowHammer bit
flips to the diminishing effects of cell-to-cell interference with distance. 

\stepcounter{obscount} \textbf{Observation \arabic{obscount}.} \emph{The number
of RowHammer bit flips that occur in a given row decreases as the distance from
the victim row increases.} 

Third, we observe that only even-numbered offsets from the victim row
contain RowHammer bit flips in all chips except LPDDR4-1x chips from
Manufacturer B.  However, the rows containing RowHammer bit flips in
Manufacturer B's LPDDR4-1x chips would be even-numbered offsets if we translate
all rows to physical rows based on our observation in
Section~\ref{subsec:methodology:characterizing_rowhammer} (i.e., divide each
row number by 2 and round down).  While we are uncertain why we observe
RowHammer bit flips only in physical even-numbered offsets from the victim row,
we believe that it may be due to the internal circuitry layout of DRAM rows.  

\begin{figure}[H] \centering
    \includegraphics[width=0.9\linewidth]{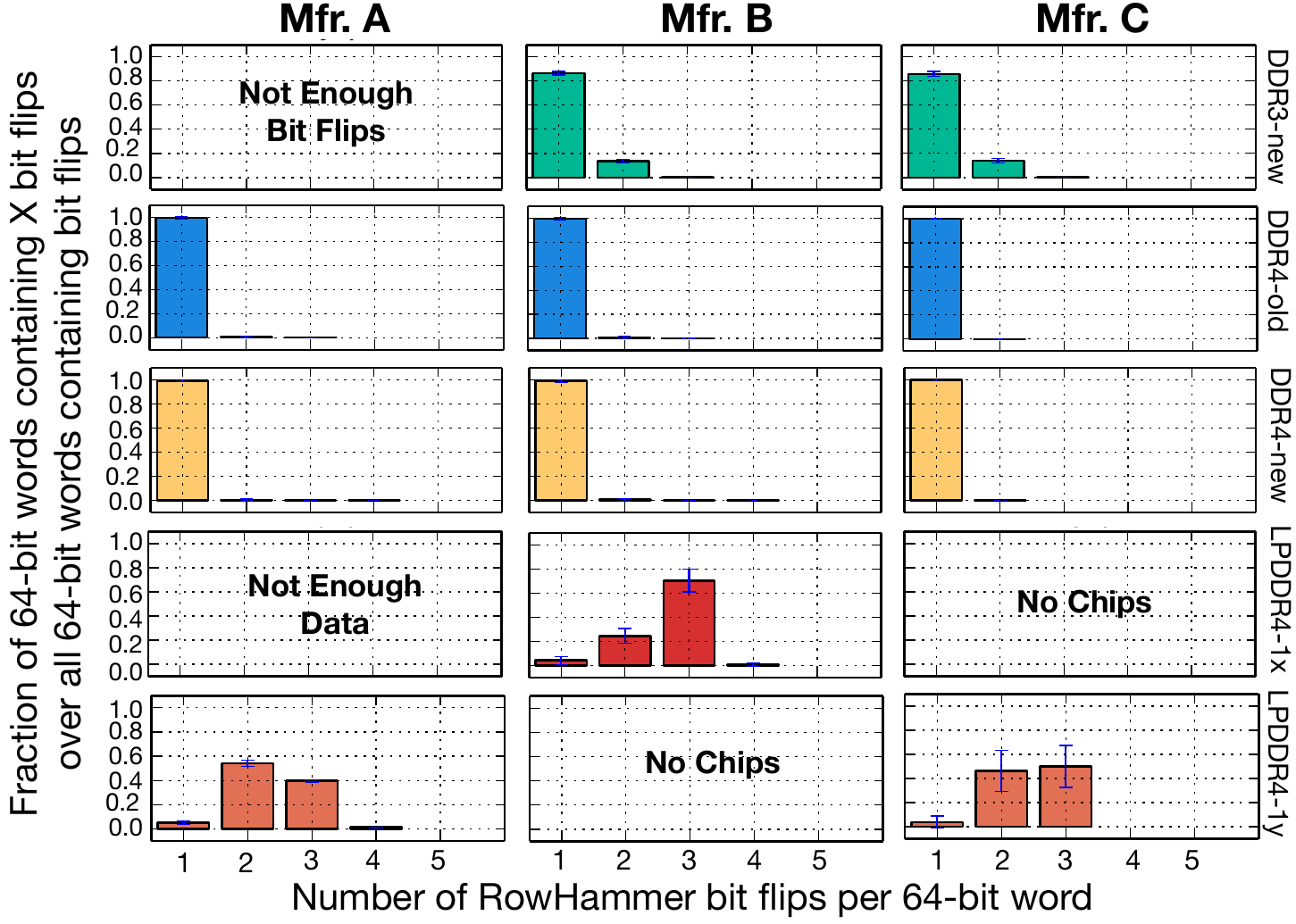}
    \caption{Distribution of the number of RowHammer bit flips per 64-bit word for each DRAM type-node configuration.} 
    \label{fig:char:density}
\end{figure}

We next study the spatial distribution of RowHammer-vulnerable DRAM cells in a
DRAM array using the same set of RowHammer bit flips.
Figure~\ref{fig:char:density} shows the distribution of 64-bit words containing
x RowHammer bit flips across our tested DRAM chips. We find the proportion of
64-bit words containing x RowHammer bit flips out of all 64-bit words in each
chip containing any RowHammer bit flip and plot the distribution as a
bar chart with error bars for each x value.

\stepcounter{obscount} \textbf{Observation \arabic{obscount}.} \emph{At a
RowHammer bit flip rate of $10^{-6}$, a single 64-bit value can contain up to
four RowHammer bit flips.} 

Because ECC~\cite{kang2014co, micron2017whitepaper, oh20153,
patel2019understanding} is typically implemented for DRAM at a 64-bit
granularity (e.g., a single-error correcting code would only protect a 64-bit
word if it contains at most one error), observation \arabic{obscount}
indicates that even at a relatively low bit flip rate of $10^{-6}$, a DRAM
chip can only be protected from RowHammer bit flips with a strong ECC code
(e.g., 4-bit error correcting code), which has high hardware overhead. 

\stepcounter{obscount} \textbf{Observation \arabic{obscount}.} \emph{The
distribution of RowHammer bit flip density per word changes significantly in
LPDDR4 chips compared to other DRAM types.} 

We find DDR3 and DDR4 chips across all manufacturers to exhibit an
exponential decay curve for increasing RowHammer bit flip densities with most
words containing only one RowHammer bit flip. However, LPDDR4 chips across
all manufacturers exhibit a much smaller fraction of words containing a
single RowHammer bit flip and significantly larger fractions of words
containing two and three RowHammer bit flips compared to DDR3 and DDR4 chips.
We believe this change in the bit flip density distribution is due to the
on-die ECC that manufacturers have included in LPDDR4 chips~\cite{kang2014co,
micron2017whitepaper, oh20153, patel2019understanding}, which is a 128-bit
single-error correcting code that corrects and hides \emph{most} single-bit
failures within a 128-bit ECC word using redundant bits (i.e.,
\emph{parity-check bits}) that are hidden from the system.  

With the failure rates at which we test, many ECC words contain
several bit flips. This exceeds the ECC's correction strength and causes the
ECC logic to behave in an undefined way. The ECC logic may 1) correct one of
the bit flips, 2) do nothing, or 3) introduce an \emph{additional} bit flip by
corrupting an error-free data bit~\cite{son2015cidra, patel2019understanding}.
On-die ECC makes single-bit errors rare because 1) any true single-bit error
is immediately corrected and 2) a multi-bit error can \emph{only} be reduced
to a single-bit error when there are no more than two bit flips within the
data bits \emph{and} the ECC logic's undefined action happens to change the
bit flip count to exactly one. In contrast, there are many more scenarios that
yield two or three bit-flips within the data bits, and a detailed experimental
analysis of how on-die ECC affects DRAM failure rates in LPDDR4 DRAM chips can
be found in~\cite{patel2019understanding}.



\subsection{First RowHammer Bit Flips} 

\begin{figure}[h] \centering
    \includegraphics[width=\linewidth]{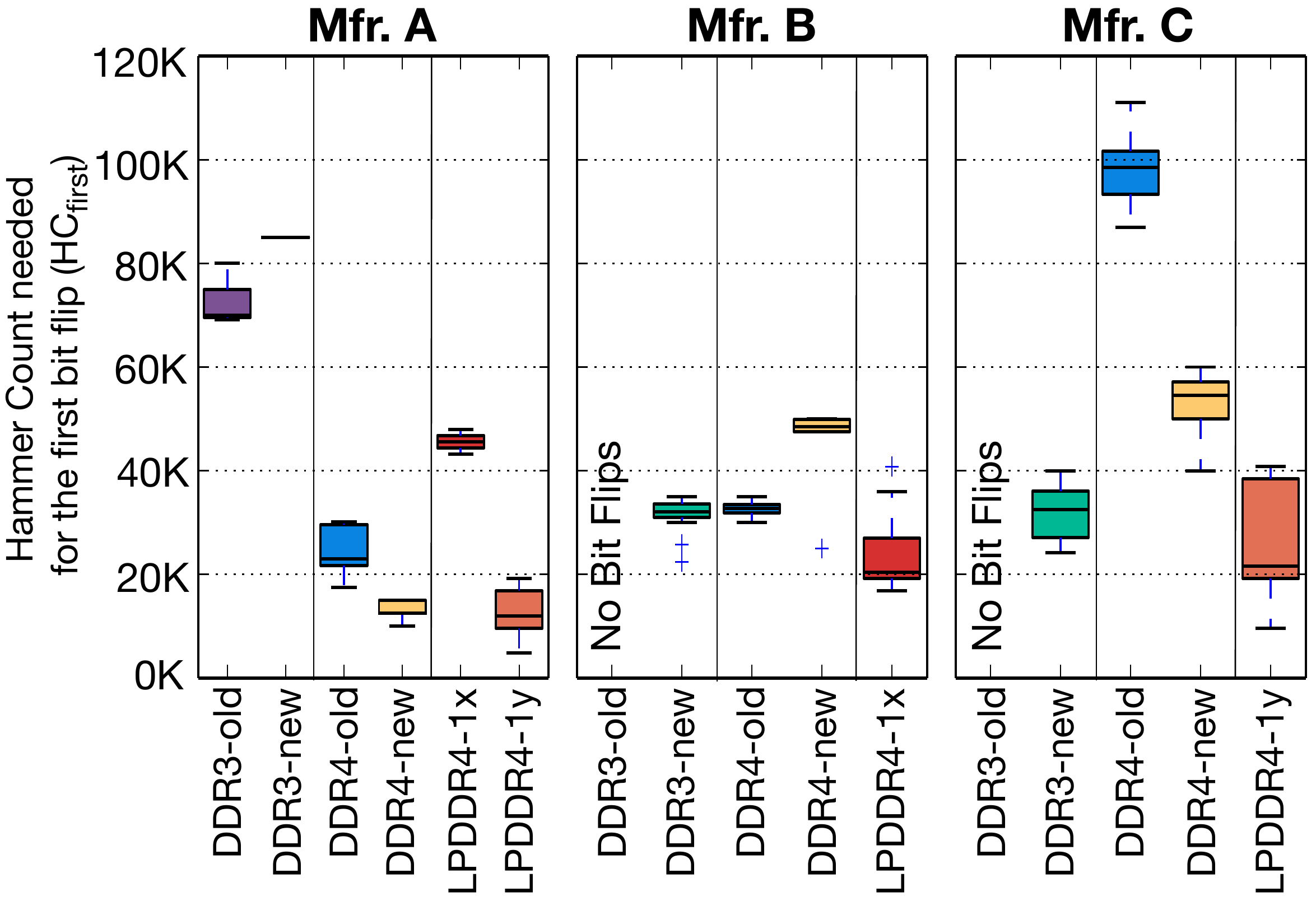}
    \caption{Number of hammers required to cause the first RowHammer bit flip ($HC_{first}$) per chip across DRAM type-node configurations.}
    \label{fig:char:first_fail}
\end{figure}

We next study the vulnerability of each chip to RowHammer. One critical
component of vulnerability to the double-sided RowHammer
attack~\cite{kim2014flipping} is identifying the weakest cell, i.e., the DRAM cell
that fails with the fewest number of accesses to physically-adjacent rows.
In order to perform this study, we sweep $HC$ at a fine granularity and record 
the $HC$ that results in the first RowHammer bit flip in the chip
({\hcfirst}).  Figure~\ref{fig:char:first_fail} plots the distribution of
{\hcfirst} across all tested chips as box-and-whisker plots.\footnote{A
box-and-whiskers plot emphasizes the important metrics of a dataset's
distribution. The box is lower-bounded by the first quartile (i.e., the median
of the first half of the ordered set of data points) and upper-bounded by the
third quartile (i.e., the median of the second half of the ordered set of data
points). The median falls within the box. The \emph{inter-quartile range} (IQR)
is the distance between the first and third quartiles (i.e., box size).
Whiskers extend an additional $1.5 \times IQR$ on either sides of the box.  We
indicate outliers, or data points outside of the range of the whiskers, with
pluses.} The subplots contain the distributions of each tested DRAM type-node 
configuration for the different DRAM manufacturers. The x-axis organizes the
distributions by DRAM type-node configuration in order of age (older on the left to younger
on the right).  We further subdivide the subplots for chips of the same DRAM
type (e.g., DDR3, DDR4, LPDDR4) with vertical lines. Chips of the same DRAM
type are colored with the same color for easier visual comparison across DRAM
manufacturers.

\stepcounter{obscount} \textbf{Observation \arabic{obscount}.} \emph{Newer
chips from a given DRAM manufacturer appear to be more vulnerable to RowHammer
bit flips. This is demonstrated by the clear reduction in {\hcfirst} 
values from old to new DRAM generations (e.g., LPDDR4-1x to LPDDR4-1y in
manufacturer A, or DDR4-old to DDR4-new in manufacturers A and C).} 

We believe this observation is due to DRAM technology process scaling
wherein both 1) DRAM cell capacitance reduces and 2) DRAM cell density
increases as technology node size reduces. Both factors together lead to more
interference between cells and likely faster charge leakage from the DRAM
cell's smaller capacitors, leading to a higher vulnerability to RowHammer. We find
two exceptions to this trend (i.e., a general increase in {\hcfirst} from
DDR3-old to DDR3-new chips of manufacturer A and from DDR4-old to DDR4-new
chips of manufacturer B), but we believe these potential
anomalies may be due to our inability to identify explicit manufacturing
dates and correctly categorize these particular chips. 

\stepcounter{obscount} \textbf{Observation \arabic{obscount}.} \emph{In
LPDDR4-1y chips from manufacturer A, there are chips whose weakest cells
fail after only 4800 hammers.} 

This observation has serious implications for the future as DRAM
technology node sizes will continue to reduce and {\hcfirst} will only get
smaller. We discuss these implications further in
Section~\ref{sec:implications}.  Table~\ref{tab:device_Nth} shows the
lowest observed {\hcfirst} value for any chip within a DRAM type-node
configuration (i.e., the minimum values of each distribution in
Figure~\ref{fig:char:first_fail}).

\begin{table}[h]
\footnotesize
\begin{center}
\renewcommand{\arraystretch}{0.9}
\renewcommand{\aboverulesep}{0ex}
\renewcommand{\belowrulesep}{0ex}
\caption{Lowest $\bm{HC_{first}}$ values ($\mathbf{\times1000}$) across all chips of each DRAM type-node configuration.} 
\begin{tabular}{l R{1.45cm} R{1.45cm} R{1.2cm}}
\toprule
\multicolumn{1}{c}{\textbf{DRAM}} & \multicolumn{3}{c}{\textbf{$\bm{HC_{first}}$ (Hammers until first bit flip) $\mathbf{\times1000}$}} \\
\multicolumn{1}{c}{\textbf{type-node}} & \quad \textbf{\textit{Mfr. A}} & \quad \textbf{\textit{Mfr. B}} & \quad \textbf{\textit{Mfr. C}} \\
\toprule
DDR3-old & 69.2 & 157 & 155 \\ 
DDR3-new & 85 & 22.4 & 24 \\
\Xhline{0.2\arrayrulewidth}
DDR4-old & 17.5 & 30 & 87  \\
DDR4-new & 10 & 25 & 40 \\
\Xhline{0.2\arrayrulewidth}
\ap{LPDDR4-1x} & 43.2 & 16.8 & N/A \\
\ap{LPDDR4-1y} & 4.8 & N/A & 9.6 \\
\toprule
\end{tabular}
\label{tab:device_Nth}
\vspace{-3mm} 
\end{center}
\end{table}
\textbf{Effects of ECC.} The use of error correcting codes (ECC) to
improve the reliability of a DRAM chip is common practice, with most
system-level~\cite{balasubramonian2019innovations, cojocar2019exploiting,
gong2018duo, kim2015bamboo} or on-die~\cite{micron2017whitepaper, kwak2017a,
kang2014co, patel2019understanding, kwon2017an} ECC mechanisms providing single
error correction capabilities at the granularity of 64- or 128-bit words.
We examine 64-bit ECCs since, for the same correction capability (e.g.,
single-error correcting), they are stronger than 128-bit ECCs. In order
to determine the efficacy with which ECC can mitigate RowHammer effects on real
DRAM chips, we carefully study three metrics across each of our chips: 1)
the lowest $HC$ required to cause the first RowHammer bit flip (i.e.,
{\hcfirst}) for a given chip (shown in Figure~\ref{fig:char:first_fail}), 2)
the lowest $HC$ required to cause at least two RowHammer bit flips (i.e.,
{\hcsecond}) within any 64-bit word, and 3) the lowest $HC$ required to cause
at least three RowHammer bit flips (i.e., {\hcthird}) within any 64-bit word.
These quantities tell us, for ECCs of varying strengths (e.g.,
single-error correction code, double-error correction code), at which
$HC$ values the ECC can 1) mitigate RowHammer bit flips and 2) no
longer reliably mitigate RowHammer bit flips for that particular chip.

\begin{figure}[h] \centering
    \includegraphics[width=0.99\linewidth]{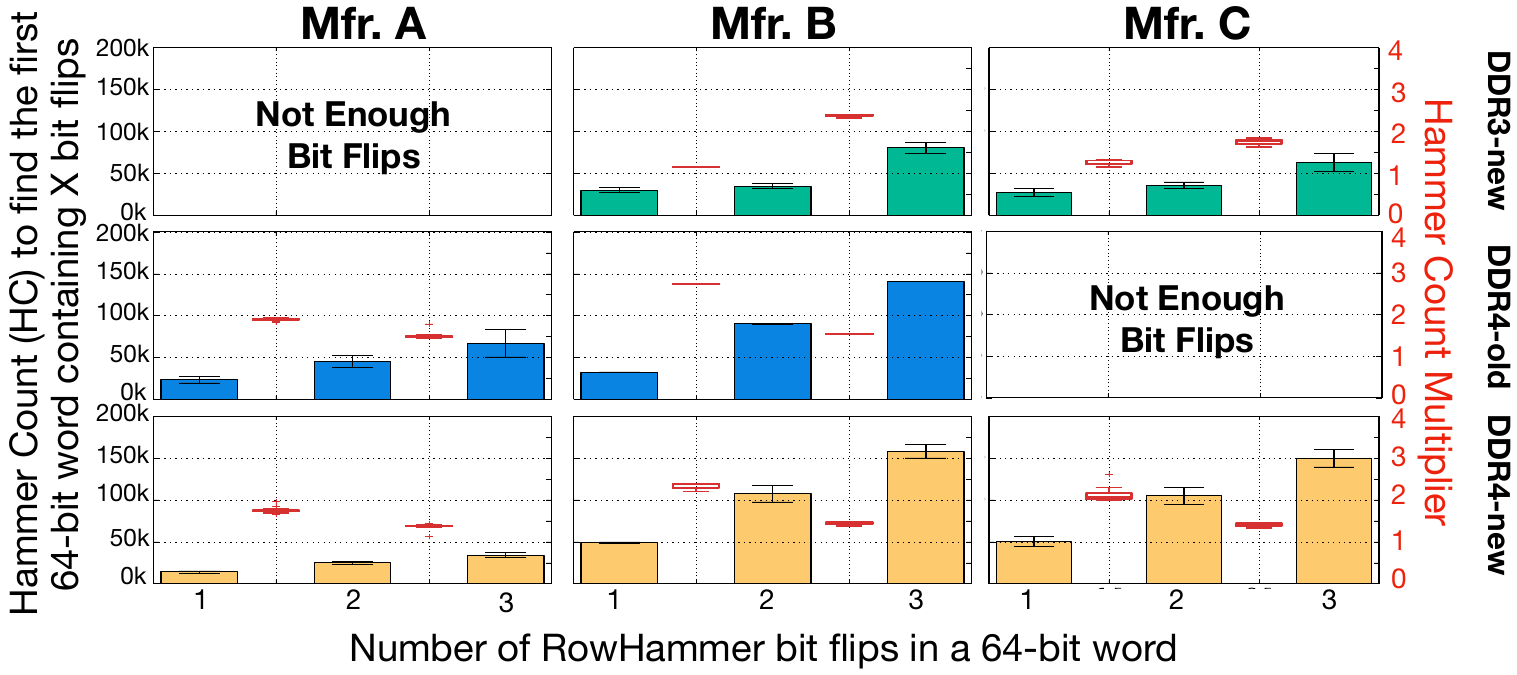}
    \caption{Hammer Count (left y-axis) required to find the first 64-bit word containing one, two, and three RowHammer bit flips. Hammer Count Multiplier (right y-axis) quantifies the $HC$ difference between every two points on the x-axis (as a multiplication factor of the left point to the right point).}
    \label{fig:ECC_efficacy} 
\end{figure}

Figure~\ref{fig:ECC_efficacy} plots as a bar graph the $HC$ (left y-axis)
required to find the first 64-bit word containing one, two, and three RowHammer
bit flips (x-axis) across each DRAM type-node configuration.  The error bars
represent the standard deviation of $HC$ values across all chips tested. On the
same figure, we also plot with red boxplots, the increase in $HC$ (right
y-axis) between the $HC$s required to find the first 64-bit word containing one
and two RowHammer bit flips, and two and three RowHammer bit flips. These
multipliers indicate how {\hcfirst} would change in a chip if the chip uses
single-error correcting ECC or moves from a single-error correcting to a
double-error correcting ECC.  Note that we 1) leave two plots (i.e., Mfr.  A
DDR3-new and Mfr. C DDR4-old) empty since we are unable to induce enough
RowHammer bit flips to find 64-bit words containing more than one bit flip in
the chips and 2) do not include data from our LPDDR4 chips because they already
include on-die ECC~\cite{micron2017whitepaper, kwak2017a, kang2014co,
patel2019understanding, kwon2017an}, which obfuscates errors potentially
exposed to any other ECC mechanisms~\cite{patel2019understanding}.  

\stepcounter{obscount} \textbf{Observation \arabic{obscount}.} \emph{A
single-error correcting code can significantly improve {\hcfirst} by up to
$2.78\times$ in DDR4-old and DDR4-new DRAM chips, and $1.65\times$ in
DDR3-new DRAM chips.} 


\stepcounter{obscount} \textbf{Observation \arabic{obscount}.}
\emph{Moving from a double-error correcting code to a triple-error
correcting code has diminishing returns in DDR4-old and DDR4-new DRAM
chips (as indicated by the reduction in the $HC$ multiplier) compared to when
moving from a single-error correcting code to a double-error
correcting code. However, using a triple-error correcting code in
DDR3-new DRAM chips continues to further improve the {\hcfirst} and thus
reduce the DRAM chips' vulnerability to RowHammer.}

\subsection{Single-Cell RowHammer Bit Flip Probability} 

We examine how the failure probability of a single RowHammer bit flip changes as $HC$
increases.  We sweep $HC$ between 25k to 150k with a step size of 5k and hammer
each DRAM row over 20 iterations. For each $HC$ value, we identify each
cell's bit flip probability (i.e., the number of times we observe a RowHammer
bit flip in that cell out of all 20 iterations). We then observe how each
cell's bit flip probability changes as $HC$ increases.  We expect that by
exacerbating the RowHammer conditions (e.g., increasing the hammer count),
the exacerbated circuit-level interference effects should result in an
increasing RowHammer bit flip probability for each individual cell.  Out of the
full set of bits that we observe \emph{any} RowHammer bit flips in,
Table~\ref{tab:sensitivity} lists the percentage of cells that have a strictly
monotonically increasing bit flip probability as we increase $HC$.

\begin{table}[h]
\footnotesize
\begin{center}
\caption{Percentage of cells with monotonically increasing RowHammer bit flip probabilities as $\bm{HC}$ increases.}
\begin{tabular}{ lR{1.35cm}R{1.35cm}R{1.2cm}}
\toprule
\multicolumn{1}{c}{\multirow{3}{*}{\begin{tabular}{c}\textbf{DRAM} \\ \textbf{type-node}\end{tabular}}} & \multicolumn{3}{c}{\textbf{Cells with monotonically increasing}} \\
 & \multicolumn{3}{c}{\textbf{ RowHammer bit flip probabilities (\%)}} \\
 & \textbf{\textit{Mfr. A}} & \textbf{\textit{Mfr. B}} & \textbf{\textit{Mfr. C}} \\
\toprule
DDR3-new & $97.6\pm0.2$ & 100 & 100 \\
\Xhline{0.2\arrayrulewidth}
DDR4-old & $98.4\pm0.1$ & 100 & 100 \\
DDR4-new & $99.6\pm0.1$ & 100 & 100 \\
\Xhline{0.2\arrayrulewidth}
LPDDR4-1x & $50.3\pm1.2$ & $52.4\pm1.4$ & N/A \\
LPDDR4-1y & $47.0\pm0.8$ & N/A & $54.3\pm5.7$ \\
\toprule
\end{tabular}
\label{tab:sensitivity}
\vspace{-3mm} 
\end{center}
\end{table}

\stepcounter{obscount} \textbf{Observation \arabic{obscount}.} \emph{For DDR3
and DDR4 chips, an overwhelming majority (i.e., more than 97\%) of the cells
tested have monotonically increasing RowHammer bit flip probabilities for
DDR3 and DDR4 chips.} 

This observation indicates that exacerbating the RowHammer
conditions by increasing $HC$ increases the probability that a DRAM cell 
experiences a RowHammer bit flip. However, we find that the proportion of cells
with monotonically increasing RowHammer bit flip probabilities as $HC$
increases is around only 50\% in the LPDDR4 chips that we test.
We believe that this decrease is due to the addition of on-die ECC in LPDDR4
chips, which can obscure the probability of observing a RowHammer bit flip from
the system's perspective in two ways. First, a RowHammer bit flip at bit X can
no longer be observable from the system's perspective if another RowHammer bit
flip at bit Y occurs within the same ECC word as a result of increasing $HC$,
and the error correction logic corrects the RowHammer bit flip at bit X.
Second, the system may temporarily observe a bit flip at bit X at
a specific $HC$ if the set of real RowHammer bit flips within an ECC word
results in a miscorrection at bit X. Since this bit flip is a result of the ECC
logic misbehaving rather than circuit-level interference, we do not observe the
expected trends for these transient miscorrected bits.

%



%% file: 6_implications_jk.tex
\section{Implications for Future Systems} 
\label{sec:implications}

Our characterization results have major implications for continued DRAM
technology scaling since DRAM's increased vulnerability to RowHammer means that
systems employing future DRAM devices will likely need to handle
significantly elevated failure rates. While prior works propose a wide variety
of RowHammer failure mitigation techniques (described in
Sections~\ref{subsec:implications:existing_mechanisms} and \ref{sec:related}),
these mechanisms will need to manage increasing failure rates going forward and
will likely suffer from high overhead (as we show in
Section~\ref{subsec:implications:evaluation}).

While DRAM and system designers currently implement several RowHammer
mitigation mechanisms (e.g., \emph{pseudo Target Row Refresh}
(pTRR)~\cite{kaczmarski2014thoughts}, \emph{Target Row Refresh}
(TRR)~\cite{lin2017handling})\footnote{Frigo et al.~\cite{frigo2020trrespass} 
recently demonstrated that these mechanisms do \emph{not} prevent \emph{all}
RowHammer bit flips from being exposed to the system, and an attacker can still
take over a system even with these mechanisms in place.}, the designers make a
number of unknown implementation choices in these RowHammer mitigation
mechanisms that are not discussed in public documentation. Therefore,
we cannot fairly evaluate how their performance overheads scale as DRAM
chips become more vulnerable to RowHammer. Instead, we evaluate five
state-of-the-art academic proposals for RowHammer mitigation
mechanisms~\cite{kim2014flipping, lee2019twice, son2017making, you2019mrloc} as
well as an ideal refresh-based mitigation mechanism.  

We evaluate each RowHammer mitigation mechanism in terms of two major
challenges that they will face going forward as they will need to support DRAM
chips more vulnerable to RowHammer: design scalability and system performance
overhead. We first qualitatively explain and discuss the five state-of-the-art
mitigation mechanisms and how they can potentially scale to support DRAM
chips that are more vulnerable to RowHammer. We then quantitatively evaluate
their performance overheads in simulation as {\hcfirst} decreases. In
order to show the opportunity for reducing performance overhead in RowHammer
mitigation, we also implement and study an \emph{ideal refresh-based mechanism}
that prevents RowHammer by refreshing a DRAM row \emph{only} immediately before
it is about to experience a bit flip. 

\subsection{RowHammer Mitigation Mechanisms}
\label{subsec:implications:existing_mechanisms}

There is a large body of work (e.g., \cite{aweke2016anvil, van2018guardion,
konoth2018zebram, brasser2016can, wu2019protecting, bock2019rip,
kim2019effective, wang2019reinforce, chakraborty2019deep, li2019detecting,
ghasempour2015armor, gruss2018another, irazoqui2016mascat}) that proposes
software-based RowHammer mitigation mechanisms. Unfortunately, many of these
works have critical weaknesses (e.g., inability to track all DRAM activations)
that make them vulnerable to carefully-crafted RowHammer attacks, as
demonstrated in some followup works (e.g., \cite{gruss2018another}).
Therefore, we focus on evaluating six mechanisms (i.e., five state-of-the-art
hardware proposals and one \emph{ideal} refresh-based mitigation
mechanism), which address a strong threat model that assumes an attacker
can cause row activations with precise memory location and timing information. We briefly
explain each mitigation mechanism and how its design scales for DRAM chips with
increased vulnerability to RowHammer (i.e., lower {\hcfirst} values). 

\textbf{Increased Refresh Rate~\cite{kim2014flipping}.} The original
RowHammer study~\cite{kim2014flipping} describes increasing the overall
DRAM refresh rate such that it is impossible to issue enough activations
within one refresh window (i.e., the time between two consecutive refresh
commands to a single DRAM row) to any single DRAM row to induce a RowHammer bit
flip. The study notes that this is an undesirable mitigation mechanism due to
its associated performance and energy overheads. In order to reliably mitigate
RowHammer bit flips with this mechanism, we scale the refresh rate such that
the refresh window (i.e., $t_{REFW}$; the time interval between consecutive
refresh commands to a single row) equals the number of hammers until the first
RowHammer bit flip (i.e., {\hcfirst}) multiplied by the activation latency
$t_{RC}$. Due to the large number of rows that must be refreshed within a
refresh window, this mechanism inherently does not scale to {\hcfirst} values
below 32k.

\textbf{PARA~\cite{kim2014flipping}.} Every time a row is opened and closed,
PARA (Probabilistic Adjacent Row Activation) refreshes one or more of the row's
adjacent rows with a low probability $p$.  Due to PARA's simple approach, it is
possible to easily tune $p$ when PARA must protect a DRAM chip with a lower
{\hcfirst} value. In our evaluation of PARA, we scale $p$ for different values
of {\hcfirst} such that the bit error rate (BER) does not exceed 1e-15 per
hour of continuous hammering.\footnote{We adopt this BER from typical consumer
memory reliability targets~\cite{micheloni2015apparatus, jedec2003failure,
patel2017reaper, cai2012error, cai2017flashtbd, luo2016enabling}.}

\textbf{ProHIT~\cite{son2017making}.} ProHIT maintains a history of DRAM
activations in a set of tables to identify any row that may be activated
{\hcfirst} times. ProHIT manages the tables probabilistically to
minimize the overhead of tracking frequently-activated DRAM rows.
ProHIT~\cite{son2017making} uses a pair of tables labeled "Hot" and "Cold" to
track the victim rows. When a row is activated, ProHIT checks whether each
adjacent row is already in either of the tables. If a \emph{row} is not in
either table, it is inserted into the cold table with a probability $p_i$. If
the table is full, the least recently inserted entry in the cold table is then
evicted with a probability $(1-p_{e}) + p_{e}/(\#cold\_entries)$ and the other
entries are evicted with a probability $p_{e}/(\#cold\_entries)$. If the row
already exists in the cold table, the row is promoted to the highest-priority
entry in the hot table with a probability $(1 - p_t) + p_t/(\#hot\_entries)$
and to other entries with a probability $p_t/(\#hot\_entries)$. If the row
already exists in the hot table, the entry is upgraded to a higher priority
position. During each refresh command, ProHIT simultaneously refreshes the row
at the top entry of the hot table, since this row has likely experienced the
most number of activations, and then removes the entry from the table. 

For ProHIT~\cite{son2017making} to effectively mitigate RowHammer with
decreasing {\hcfirst} values, the size of the tables and the probabilities for
managing the tables (e.g., $p_i$, $p_e$, $p_t$) must be adjusted. Even though
Son et al. show a low-cost mitigation mechanism for a specific {\hcfirst} value
(i.e., 2000), they do \emph{not} provide models for appropriately setting these
values for arbitrary {\hcfirst} values and how to do so is not intuitive.
Therefore, we evaluate ProHIT only when {\hcfirst}$\,= 2000$. 

\textbf{MRLoc}~\cite{you2019mrloc}. MRLoc refreshes a victim row using a
probability that is dynamically adjusted based on each row's access history.
This way, according to memory access locality, the rows that have been recorded
as a victim more recently have a higher chance of being refreshed. MRLoc uses a
queue to store victim row addresses on each activation. Depending on the time
between two insertions of a given victim row into the queue, MRLoc adjusts the
probability with which it issues a refresh to the victim row that is present in
the queue.

MRLoc's parameters (the queue size and the parameters used to calculate the
probability of refresh) are tuned for {\hcfirst}$\,= 2000$. You et
al.~\cite{you2019mrloc} choose the values for these parameters
empirically, and there is no concrete discussion on how to adjust these
parameters as {\hcfirst} changes.  Therefore we evaluate MRLoc for only
{\hcfirst}$\,= 2000$. 

As such, even though we quantitatively evaluate both
ProHIT~\cite{son2017making} and MRLoc~\cite{you2019mrloc} for completeness and
they may seem to have good overhead results at one data point, we are unable to
demonstrate how their overheads scale as DRAM chips become more vulnerable to
RowHammer. 

\textbf{TWiCe}~\cite{lee2019twice}. TWiCe tracks the number of times a
victim row's aggressor rows are activated using a table of counters and
refreshes a victim row when its count is above a threshold such that RowHammer
bit flips cannot occur. TWiCe uses two counters per entry: 1) a lifetime
counter, which tracks the length of time the entry has been in the table, and
2) an \emph{activation counter}, which tracks the number of times an aggressor
row is activated. The key idea is that TWiCe can use these two counters to
determine the rate at which a row is being hammered and can quickly prune
entries that have a low rate of being hammered. TWiCe also minimizes its table
size based on the observation that the number of rows that can be
activated enough times to induce RowHammer failures within a refresh window is
bound by the DRAM chip's vulnerability to RowHammer. 

When a row is activated, TWiCe checks whether its adjacent rows are already in
the table. If so, the activation count for each row is incremented. Otherwise,
new entries are allocated in the table for each row. Whenever a row's
activation count surpasses a threshold $t_{RH}$ defined as {\hcfirst}$/4$,
TWiCe refreshes the row. TWiCe also defines a pruning stage that 1) increments
each lifetime counter, 2) checks each row's hammer rate based on both counters,
and 3) prunes entries that have a lifetime hammer rate lower than a
\emph{pruning threshold}, which is defined as $t_{RH}$ divided by the
number of refresh operations per refresh window (i.e.,
$t_{RH}/(t_{REFW}/t_{REFI})$). TWiCe performs pruning operations during refresh
commands so that the latency of a pruning operation is hidden behind the DRAM
refresh commands.

If $t_{RH}$ is lower than the number of refresh intervals in a refresh window
(i.e., $8192$), a couple of complications arise in the design. TWiCe either 1) cannot
prune its table, resulting in a very large table size since every row that is
accessed at least once will remain in the table until the end of the refresh
window or 2) requires floating point operations in order to calculate
thresholds for pruning, which would significantly increase the latency of the
pruning stage.  Either way, the pruning stage latency would increase
significantly since a larger table also requires more time to check each entry,
and the latency may no longer be hidden by the refresh command. 

As a consequence, TWiCe does \emph{not} support $t_{RH}$ values lower than the
number of refresh intervals in a refresh window ($\sim8k$ in several DRAM
standards, e.g., DDR3, DDR4, LPDDR4). This means that in its current form, we
\emph{cannot} fairly evaluate TWiCe for {\hcfirst} values below $32k$, as
$t_{RH} =\,${\hcfirst}$/4$. However, we do evaluate an ideal version of
TWiCe (i.e., \emph{TWiCe-ideal}) for {\hcfirst} values below $32k$ assuming
that TWiCe-ideal solves \emph{both} issues of the large table size and the
high-latency pruning stage at lower {\hcfirst} values. 

\textbf{Ideal Refresh-based Mitigation Mechanism.} We implement an ideal
refresh-based mitigation mechanism that tracks all activations to every row in
DRAM and issues a refresh command to a row only right before it can potentially 
experience a RowHammer bit flip (i.e., when a physically-adjacent row has
been activated {\hcfirst} times). 

\subsection{Evaluation of Viable Mitigation Mechanisms}
\label{subsec:implications:evaluation}

We first describe our methodology for evaluating the five state-of-the-art
RowHammer mitigation mechanisms (i.e., increased refresh
rate~\cite{kim2014flipping}, PARA~\cite{kim2014flipping},
ProHIT~\cite{son2017making}, MRLoc~\cite{you2019mrloc},
TWiCe~\cite{lee2019twice}) and the ideal refresh-based mitigation mechanism.

\subsubsection{Evaluation Methodology}~
\label{subsec:implications:methodology}
We use Ramulator~\cite{ramulatorgithub, kim2016ramulator}, a cycle-accurate
DRAM simulator with a simple core model and a system configuration as listed in
Table~\ref{tab:ramulator_config}, to implement and evaluate the RowHammer
mitigation mechanisms. To demonstrate how the performance overhead of each
mechanism would scale to future devices, we implement, to the best of our
ability, parameterizable methods for scaling the mitigation mechanisms to DRAM
chips with varying degrees of vulnerability to RowHammer (as described in
Section~\ref{subsec:implications:existing_mechanisms}). 

\begin{table}[ht]
	\setlength\tabcolsep{1.5pt} 
    \footnotesize 
    \centering
    \caption{System configuration for simulations.} 
    \begin{tabular}{m{2.5cm}m{5.8cm}}
    \toprule
    \textbf{Parameter}     & \textbf{Configuration}  \\\toprule 
    Processor          & 4GHz, 8-core, 4-wide issue, 128-entry instr. window \\\hline
    Last-level Cache   & 64-Byte cache line, 8-way set-associative, 16MB \\\hline
    Memory Controller  & 64 read/write request queue, FR-FCFS~\cite{rixner2000memory, zuravleff1997controller} \\\hline
	Main Memory        & DDR4, 1-channel, 1-rank, 4-bank groups, 4-banks per bank group, 16k rows per bank\\ \toprule
    \end{tabular}
    \label{tab:ramulator_config}
	\vspace{-3mm} 
\end{table}

\textbf{Workloads.} We evaluate 48 8-core workload mixes drawn randomly from
the full SPEC CPU2006 benchmark suite~\cite{spec2006} to demonstrate the
effects of the RowHammer mitigation mechanisms on systems during typical use
(and \emph{not} when a RowHammer attack is being mounted).  The set of
workloads exhibit a wide range of memory intensities. The workloads' MPKI
values (i.e., last-level cache misses per kilo-instruction) range from 10 to
740. This wide range enables us to study the effects of RowHammer mitigation
on workloads with widely varying degrees of memory intensity. We note that there could
be other workloads with which mitigation mechanisms exhibit higher performance
overheads, but we did not try to maximize the overhead experienced by workloads
by biasing the workload construction in any way. 
We simulate each workload until each core executes at least 200 million
instructions. For all configurations, we initially warm up the caches by
fast-forwarding 100 million instructions.

\textbf{Metrics.} Because state-of-the-art RowHammer mitigation
mechanisms rely on additional DRAM refresh operations to prevent RowHammer, we
use two different metrics to evaluate their impact on system performance.
First, we measure \emph{DRAM bandwidth overhead}, which quantifies the
fraction of the total system DRAM bandwidth consumption coming from the
RowHammer mitigation mechanism. Second, we measure overall workload performance
using the \emph{weighted speedup} metric~\cite{eyerman2008system,
snavely2000symbiotic}, which effectively measures job throughput for multi-core
workloads~\cite{eyerman2008system}. We normalize the weighted speedup to
its baseline value, which we denote as 100\%, and find that when using
RowHammer mitigation mechanisms, most values appear below the baseline.
Therefore, for clarity, we refer to normalized weighted speedup as
\emph{normalized system performance} in our evaluations.



\subsubsection{Evaluation of Mitigation Mechanisms}
\begin{figure}\centering
    \includegraphics[width=0.95\linewidth]{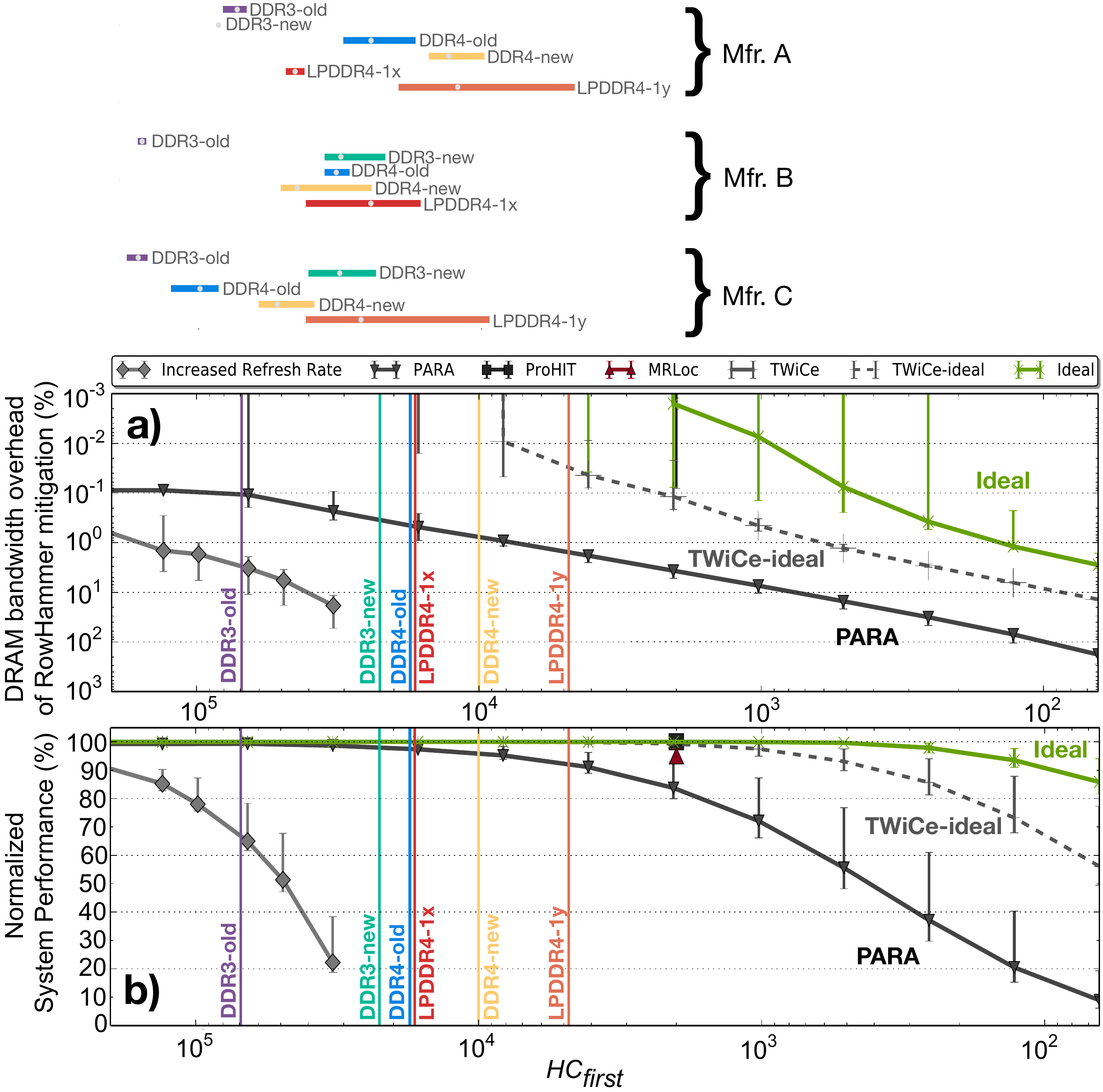} 
    \caption{Effect of RowHammer mitigation mechanisms on a) DRAM bandwidth overhead (note the inverted log-scale y-axis) and b) system performance, as DRAM chips become more vulnerable to RowHammer (from left to right).} 
    \label{fig:rh_mitigation_overhead_perf}
	\vspace{-2mm} 
\end{figure}

~
Figure~\ref{fig:rh_mitigation_overhead_perf} shows the results of our
evaluation of the RowHammer mitigation mechanisms (as described in
Section~\ref{subsec:implications:existing_mechanisms}) for chips of varying
degrees of RowHammer vulnerability (i.e., $200k \geq\,${\hcfirst}$\,\geq 64$)
for our two metrics: 1) DRAM bandwidth overhead in
Figure~\ref{fig:rh_mitigation_overhead_perf}a and 2) normalized system
performance in Figure~\ref{fig:rh_mitigation_overhead_perf}b. Each data
point shows the average value across 48 workloads with minimum and maximum
values drawn as error bars. 

For each DRAM type-node configuration that we characterize, we plot the minimum
{\hcfirst} value found across chips within the configuration (from
Table~\ref{tab:device_Nth}) as a vertical line to show how each RowHammer
mitigation mechanism would impact the overall system when using a DRAM chip of
a particular configuration. Above the figures (sharing the x-axis with
Figure~\ref{fig:rh_mitigation_overhead_perf}), we draw horizontal lines
representing the ranges of {\hcfirst} values that we observe for every tested
DRAM chip per DRAM type-node configuration across manufacturers. We color the
ranges according to DRAM type-node configuration colors in the figure, and
indicate the average value with a gray point.  Note that these lines directly
correspond to the box-and-whisker plot ranges in
Figure~\ref{fig:char:first_fail}. 

We make \emph{five} key observations from this figure. First, DRAM 
bandwidth overhead is highly correlated with normalized system performance,
as DRAM bandwidth consumption is the main source of system interference caused by 
RowHammer mitigation mechanisms. We note that several points (i.e., 
ProHIT, MRLoc, and TWiCe and Ideal evaluated at higher {\hcfirst} values) are not visible
in Figure~\ref{fig:rh_mitigation_overhead_perf}a since we are plotting an
inverted log graph and these points are very close to zero.  Second, in the
latest DRAM chips (i.e., the LPDDR4-1y chips), only PARA, ProHIT, and MRLoc 
are viable options for mitigating RowHammer bit flips with reasonable average
normalized system performance: 92\%, 100\%, and 100\%, respectively. 
Increased Refresh Rate and TWiCe do not scale to such degrees of RowHammer vulnerability
(i.e., {\hcfirst}$\,= 4.8k$), as discussed in
Section~\ref{subsec:implications:existing_mechanisms}.  Third, only PARA's
design scales to low {\hcfirst} values that we may see in future DRAM chips,
but has very low average normalized system performance (e.g., 72\% when
{\hcfirst}$\,=1024$; 47\% when {\hcfirst}$\,=256$; 20\% when {\hcfirst}$\,=128$).
While TWiCe-ideal has higher normalized system performance over PARA (e.g.,
98\% when {\hcfirst}$\,=1024$; 86\% when {\hcfirst}$\,=256$; 73\% when
{\hcfirst}$\,=128$), there are significant practical limitations in enabling
TWiCe-ideal for such low {\hcfirst} values (discussed in
Section~\ref{subsec:implications:existing_mechanisms}). Fourth, 
ProHIT and MRLoc both exhibit high normalized system performance at their single data
point (i.e., 95\% and 100\%, respectively when {\hcfirst}$\,= 2000$), but
these works do not provide models for scaling their mechanisms to lower
{\hcfirst} values and how to do so is not intuitive (as described in
Section~\ref{subsec:implications:existing_mechanisms}). Fifth, the ideal
refresh-based mitigation mechanism is \emph{significantly} and increasingly
better than any existing mechanism as {\hcfirst} reduces below $1024$.
This indicates that there is still significant opportunity for developing a
refresh-based RowHammer mitigation mechanism with low performance overhead that
scales to low {\hcfirst} values.  However, the ideal mechanism affects system
performance at very low {\hcfirst} values (e.g., 99.96\% when
{\hcfirst}$\,=1024$; 97.91\% when {\hcfirst}$\,=256$; 93.53\% when
{\hcfirst}$\,=128$), indicating the potential need for a better approach to
solving RowHammer in future ultra-dense DRAM chips. 

We conclude that while existing mitigation mechanisms may exhibit
reasonably small performance overheads for mitigating RowHammer bit flips in
modern DRAM chips, their overheads do \emph{not} scale well in future DRAM
chips that will likely exhibit higher vulnerability to RowHammer. Thus, we need
new mechanisms and approaches to RowHammer mitigation that will scale to DRAM
chips that are highly vulnerable to RowHammer bit flips.





\subsection{RowHammer Mitigation Going Forward}
\label{sec:profiling}


DRAM manufacturers continue to adopt smaller technology nodes to improve DRAM
storage density and are forecasted to reach 1z and 1a technology nodes within
the next couple of years~\cite{DRAMtechroadmap}. Unfortunately, our findings
show that future DRAM chips will likely be increasingly vulnerable to
RowHammer. This means that, to maintain market competitiveness without
suffering factory yield loss, manufacturers will need to develop
effective RowHammer mitigations for coping with increasingly
vulnerable DRAM chips.

\subsubsection{Future Directions in RowHammer Mitigation} 
\label{subsec:implications:mitigation_solutions} 

~RowHammer mitigation mechanisms have been proposed across the computing
stack ranging from circuit-level mechanisms built into the DRAM chip itself to
system-level mechanisms that are agnostic to the particular DRAM chip that the
system uses.  Of these solutions, our evaluations in
Section~\ref{subsec:implications:existing_mechanisms} show that, while the
ideal refresh-based RowHammer mitigation mechanism, which inserts the
minimum possible number of additional refreshes to prevent RowHammer bit flips,
scales reasonably well to very low {\hcfirst} values (e.g., only 6\%
performance loss when {\hcfirst} is 128), existing RowHammer mitigation
mechanisms either \emph{cannot} scale or cause severe system performance
penalties when they scale. 






To develop a scalable and low-overhead mechanism that can prevent
RowHammer bit flips in DRAM chips with a high degree of RowHammer vulnerability
(i.e., with a low {\hcfirst} value), we believe it is essential to explore
all possible avenues for RowHammer mitigation. Going forward, we identify two
promising research directions that can potentially lead to new
RowHammer solutions that can reach or exceed the scalability of the
ideal refresh-based mitigation mechanism: (1) DRAM-system cooperation and (2)
profile-guided mechanisms. The remainder of this section briefly discusses
our vision for each of these directions.

\noindent
\textbf{DRAM-System Cooperation.} Considering either DRAM-based or
system-level mechanisms alone ignores the potential benefits of addressing the 
RowHammer vulnerability from both perspectives together. While the root causes
of RowHammer bit flips lie within DRAM, their negative effects are observed at the
system-level.  Prior work~\cite{mutlu2014research, mutlu2013memory} stresses
the importance of tackling these challenges at all levels of the stack, and we
believe that a holistic solution can achieve a high degree of protection at
relatively low cost compared to solutions contained within either domain alone.

\noindent
\textbf{Profile-Guided Mechanisms.} The ability to accurately profile for
RowHammer-susceptible DRAM cells or memory regions can provide a powerful
substrate for building targeted RowHammer solutions that efficiently
mitigate RowHammer bit flips at low cost. Knowing (or effectively predicting)
the locations of bit flips before they occur in practice could lead to a large
reduction in RowHammer mitigation overhead, providing new information that no
known RowHammer mitigation mechanism exploits today. For example, within the
scope of known RowHammer mitigation solutions, increasing the refresh rate can
be made far cheaper by only increasing the refresh rate for known-vulnerable
DRAM rows. Similarly, ECC or DRAM access counters can be used only for
known-vulnerable cells, and even a software-based mechanism can be adapted
to target only known-vulnerable rows (e.g., by disabling them or remapping
them to reliable memory).

Unfortunately, there exists no such effective RowHammer error profiling
methodology today. Our characterization in this work essentially follows the
na{\"i}ve approach of individually testing each row by attempting to induce
the worst-case testing conditions (e.g., $HC$, data pattern, ambient
temperature etc.). However, this approach is extremely time consuming due to
having to test each row individually (potentially multiple times with various 
testing conditions). Even for a relatively small DRAM module of 8GB with 8KB
rows, hammering each row only once for only one refresh window of 64ms requires
over 17 hours of continuous testing, which means that the na{\"i}ve approach to
profiling is infeasible for a general mechanism that may be used in a
production environment or for online operation. We believe that developing a
fast and effective RowHammer profiling mechanism is a key research challenge,
and we hope that future work will use the observations made in this study
and other RowHammer characterization studies to find a solution.

%% file: 7_related.tex
\section{Related Work}
\label{sec:related}


Although many works propose RowHammer attacks and mitigation mechanisms, only
three works~\cite{kim2014flipping, park2016statistical, park2016experiments}
provide detailed failure-characterization studies that examine how RowHammer
failures manifest in real DRAM chips. However, none of these studies
show how the number of activations to induce RowHammer bit flips is
changing across modern DRAM types and generations, and the original RowHammer
study~\cite{kim2014flipping} is already six years old and limited to DDR3
DRAM chips only. This section highlights the most closely related prior works
that study the RowHammer vulnerability of older generation chips or examine
other aspects of RowHammer.

\textbf{Real Chip Studies.} Three key studies (i.e., the pioneering
RowHammer study~\cite{kim2014flipping} and two subsequent
studies~\cite{park2016statistical, park2016experiments}) perform extensive
experimental RowHammer failure characterization using older DDR3 devices.
However, these studies are restricted to only DDR3 devices and do not provide
a scaling study of hammer counts across DRAM types and generations. In
contrast, our work provides the first rigorous experimental study showing how
RowHammer characteristics scale across different DRAM generations and how DRAM
chips designed with newer technology nodes are increasingly vulnerable to
RowHammer. Our work complements and furthers the analyses provided in prior
studies.


\textbf{Simulation Studies.} Yang et al.~\cite{yang2019trap} use device-level
simulations to explore the root cause of the RowHammer vulnerability. While
their analysis identifies a likely explanation for the failure mechanism
responsible for RowHammer, they do not present experimental data taken from
real devices to support their conclusions.






\textbf{RowHammer Mitigation Mechanisms.} Many prior works~\cite{you2019mrloc,
son2017making, aweke2016anvil, konoth2018zebram, van2018guardion,
brasser2016can, kim2014flipping, kim2014architectural, irazoqui2016mascat,
gomez2016dram, lee2019twice, bu2018srasa, bains2015rowref, bains14d, bains14c,
greenfield14b, bains2016row, bains2015row, rh-apple, rh-hp, rh-lenovo,
rh-cisco, hassan2019crow, wu2019protecting, bock2019rip, kim2019effective,
wang2019reinforce, fisch2017dram, chakraborty2019deep, li2019detecting,
wang2019detect} propose RowHammer mitigation techniques. Additionally, several
patents for RowHammer prevention mechanisms have been filed~\cite{bains2015row,
bains14d, bains14c, bains2016row, bains2015rowref, greenfield14a}. However,
these works do not analyze how their solutions will scale to future DRAM
generations and do not provide detailed failure characterization data from
modern DRAM devices. Similar and other related works on RowHammer can be found
in a recent retrospective~\cite{mutlu2019rowhammer}.


%% file: 8_conclusion.tex
\section{Conclusion} 

We provide the first rigorous experimental RowHammer failure characterization
study that demonstrates how the RowHammer vulnerability of modern DDR3, DDR4,
and LPDDR4 DRAM chips scales across DRAM generations and technology nodes.
Using experimental data from 1580 real DRAM chips produced by the three
major DRAM manufacturers, we show that modern DRAM chips that use smaller
process technology node sizes are significantly more vulnerable to RowHammer
than older chips. Using simulation, we show that existing RowHammer mitigation 
mechanisms 1) suffer from prohibitively large performance overheads at
projected future hammer counts and 2) are still far from an \emph{ideal}
selective-refresh-based RowHammer mitigation mechanism. Based on our study, we
motivate the need for a scalable and low-overhead solution to RowHammer and
provide two promising research directions to this end. We hope that the results
of our study will inspire and aid future work to develop efficient solutions
for the RowHammer bit flip rates we are likely to see in DRAM chips in the near
future.

%% file: appendix.tex
\appendix

\section{Appendix Tables}

\input{chip_table_DDR3} 
\input{chip_table}

%% file: chip_table_DDR3.tex
\begin{table}[H]
  \centering
  \caption{\parbox{\textwidth}{Sample population of 60 DDR3 DRAM modules, categorized by manufacturer and sorted by manufacturing date.}}
    \label{table:ddr3_table}%
\end{table}%

\begin{center}
\vspace{-4ex}
  \resizebox{\textwidth}{!}{
    \begin{tabular}{lllccccccc}
        \toprule
        \mr{3.5}{Manufacturer} & \mr{3.5}{Tech. Node} & \mr{3.5}{\emph{Module}} & \mr{3.5}{\emph{\makecell{Date\\(yy-ww)}}} &
        \multicolumn{2}{c}{\emph{Timing}} & \multicolumn{3}{c}{\emph{Organization}} & \mr{3.5}{\emph{Minimum $HC_{first}$}}  \\
        \cmidrule(lr){5-6}\cmidrule(lr){7-9}
        \multicolumn{1}{c}{} & Generation & \multicolumn{2}{c}{} & \emph{\makecell{Freq.\\(MT/s)}} & \emph{\makecell{tRC\\(ns)}} &
        \emph{\makecell{Size\\(GB)}} & \emph{Chips} & \emph{Pins} & \emph{(x1000)} \\
        \midrule

        \stripe
	\cellcolor{white} & \cellcolor{white} & $A_{0}$     & 10-19 & 1066 & 50.625 & 1 & 8 & x8 & 155 \\ 
	\cellcolor{white} & \cellcolor{white} & $A_1$       & 10-40 & 1333 & 49.5 & 2 & 8 & x8 & N/A \\ 
        \stripe
	\cellcolor{white} & \cellcolor{white} & $A_{2-6}$   & 12-11 & 1866 & 47.91 & 2 & 8 & x8 & 156 \\ 
	\cellcolor{white} & \cellcolor{white}{\multirow{-4}{5em}{old}} & $A_{7-9}$   & 12-32 & 1600 & 48.75 & 2 & 8 & x8 & 69.2 \\ 
		\cmidrule(lr){2-10} 
        \stripe
	\cellcolor{white} & \cellcolor{white} & $A_{10-16}$ & 14-16 & 1600 & 48.75 & 4 & 8 & x8 & 85 \\ 
	\cellcolor{white} & \cellcolor{white} & $A_{17-18}$ & 14-26 & 1600 & 48.75 & 2 & 4 & x16 & 160 \\ 
        \stripe
	\cellcolor{white} & \cellcolor{white}{\multirow{-3}{5em}{new}} & $A_{19}$    & 15-23 & 1600 & 48.75 & 8 & 16 & x4 & 155 \\ 
    \midrule 
		\multirow{-10}{5em}{\centering {\LARGE A}} 
	\cellcolor{white} & \cellcolor{white} & $B_{0-1}$   & 10-48 & 1333 & 49.5 & 1 & 8 & x8 & N/A \\
        \stripe
	\cellcolor{white} & \cellcolor{white} & $B_{2-4}$   & 11-42 & 1333 & 49.5 & 2 & 8 & x8 & N/A \\ 
	\cellcolor{white} & \cellcolor{white} & $B_{5-6}$   & 12-24 & 1600 & 48.75 & 2 & 8 & x8 & 157 \\ 
        \stripe
	\cellcolor{white} & \cellcolor{white}{\multirow{-4}{5em}{old}} & $B_{7-10}$  & 13-51 & 1600 & 48.75 & 4 & 8 & x8 & N/A \\ 
		\cmidrule(lr){2-10} 
	\cellcolor{white} & \cellcolor{white} & $B_{11-14}$ & 15-22 & 1600 & 50.625 & 4 & 8 & x8 & 33.5 \\ 
        \stripe
	\cellcolor{white} & \cellcolor{white}{\multirow{-2}{5em}{new}} & $B_{15-19}$ & 15-25 & 1600 & 48.75 & 2 & 4 & x16 & 22.4 \\ 
    \midrule 
		\multirow{-9}{5em}{\centering {\LARGE B}} 
	\cellcolor{white} & \cellcolor{white}{\multirow{1}{5em}{old}} & $C_{0-6}$ & 10-43 & 1333 & 49.125 & 1 & 4 & x16 & 155 \\ 
        \stripe
		\cmidrule(lr){2-10} 
        \stripe
	\cellcolor{white} & \cellcolor{white} & $C_{7}$     & 15-04 & 1600 & 48.75 & 4 & 8 & x8 & N/A \\ 
	\cellcolor{white} & \cellcolor{white} & $C_{8-12}$  & 15-46 & 1600 & 48.75 & 2 & 8 & x8 & 33.5 \\ 
        \stripe
		\multirow{-4}{5em}{\centering {\LARGE C}} 
	\cellcolor{white} & \cellcolor{white}{\multirow{-3}{5em}{new}} & $C_{13-19}$ & 17-03 & 1600 & 48.75 & 4 & 8 & x8 & 24 \\ 
    \midrule 
    \end{tabular}%
    } 
\end{center}

%% file: chip_table.tex
\begin{table}[H]
  \centering
  \caption{\parbox{\textwidth}{Sample population of 110 DDR4 DRAM modules, categorized by manufacturer and sorted by manufacturing date.}}
  \label{table:ddr4_table}%
\end{table}%

\begin{center}
\vspace{-3ex}
  \resizebox{\textwidth}{!}{
    \begin{tabular}{lllccccccc}
        \toprule
        \mr{3.5}{Manufacturer} & \mr{3.5}{Tech. Node} & \mr{3.5}{\emph{Module}} & \mr{3.5}{\emph{\makecell{Date\\(yy-ww)}}} &
        \multicolumn{2}{c}{\emph{Timing}} & \multicolumn{3}{c}{\emph{Organization}} & \mr{3.5}{\emph{Minimum $HC_{first}$}}  \\
        \cmidrule(lr){5-6}\cmidrule(lr){7-9}
        \multicolumn{1}{c}{} & Generation & \multicolumn{2}{c}{} & \emph{\makecell{Freq.\\(MT/s)}} & \emph{\makecell{tRC\\(ns)}} &
        \emph{\makecell{Size\\(GB)}} & \emph{Chips} & \emph{Pins} &  \emph{(x1000)} \\
        \midrule

        \stripe
        \cellcolor{white} & \cellcolor{white}{old}  & $A_{0-15}$    & 17-08 & 2133  & 47.06 & 4     & 8     & x8    & 17.5 \\ 
		\cmidrule(lr){2-10} 

        \cellcolor{white} & \cellcolor{white} & $A_{16-18}$   & 19-19 & 2400  & 46.16 & 4     & 4     & x16   & 12.5 \\
        \stripe
        \cellcolor{white} & \cellcolor{white} & $A_{19-24}$   & 19-36 & 2666  & 46.25 & 4     & 4     & x16   & 10 \\
        \cellcolor{white} & \cellcolor{white} & $A_{25-33}$   & 19-45 & 2666  & 46.25 & 4     & 4     & x16   & 10 \\
        \stripe
        \cellcolor{white} & \cellcolor{white} & $A_{34-36}$   & 19-51 & 2133  & 46.5  & 8     & 8     & x8    & 10 \\
        \cellcolor{white} & \multirow{-4}{5em}{new} & $A_{37-46}$   & 20-07 & 2400  & 46.16 & 8     & 8     & x8    & 12.5 \\
        \stripe
        \cellcolor{white} & \cellcolor{white} & $A_{47-58}$   & 20-08 & 2133  & 46.5  & 4     & 8     & x8    & 10 \\

		\midrule 
		\multirow{-10}{5em}{\centering {\LARGE A}} 
		\cellcolor{white} & \cellcolor{white} old & $B_{0-2}$     & N/A   & 2133  & 46.5  & 4     & 8     & x8    & 30 \\ 
		\cmidrule(lr){2-10} 
        \stripe
		\cellcolor{white} & \cellcolor{white} new & $B_{3-4}$     & N/A   & 2133  & 46.5  & 4     & 8     & x8    & 25 \\ 
		\midrule 
		\multirow{-5}{5em}{\centering {\LARGE B}} 
        \cellcolor{white} & \cellcolor{white} & $C_{0-7}$     & 16-48 & 2133  & 46.5  & 4     & 8     & x8    & 147.5\\
        \stripe
        \cellcolor{white} & \cellcolor{white} & $C_{8-17}$    & 17-12 & 2133  & 46.5  & 4     & 8     & x8    & 87 \\
		\cmidrule(lr){2-10} 

        \cellcolor{white} & \multirow{-5}{5em}{old} & $C_{45}$      & 19-01 & 2400  & 45.75 & 8     & 8     & x8    & 54 \\
        \stripe
        \cellcolor{white} & \cellcolor{white} & $C_{44}$      & 19-06 & 2400  & 45.75 & 8     & 8     & x8    & 63 \\
        \cellcolor{white} & \cellcolor{white} & $C_{34}$      & 19-11 & 2400  & 45.75 & 4     & 4     & x16   & 62.5 \\
        \stripe
        \cellcolor{white} & \cellcolor{white} & $C_{35-36}$   & 19-23 & 2400  & 45.75 & 4     & 4     & x16   & 63 \\
        \cellcolor{white} & \cellcolor{white} & $C_{37-43}$   & 19-44 & 2133  & 46.5  & 8     & 8     & x8    & 57.5 \\
        \stripe
        \cellcolor{white} & \cellcolor{white}{\multirow{-5}{5em}{new}} & $C_{18-27}$   & 19-48 & 2400  & 45.75 & 8     & 8     & x8    & 52.5 \\
		\multirow{-8}{5em}{\centering {\LARGE C}} 
		\cellcolor{white} & \cellcolor{white} & $C_{28-33}$   & N/A   & 2666  & 46.5  & 4     & 8     & x4    & 40 \\ 
    \midrule 
    \end{tabular}%
    } 

\end{center}